\documentclass{nsr}

\usepackage{amsmath,graphicx,array}
\usepackage{dcolumn,soul}%

\usepackage{amsthm}
\usepackage[figuresright]{rotating}%
\usepackage{algorithm, algorithmicx, algpseudocode}
\usepackage{listings}%
\usepackage{hyperref}
\usepackage{booktabs}
\makeatletter
\def\uns{\ifmmode\,\else$\,$\fi}%

\makeatother
 
\jvol{XX}
\jnum{X}
\jyear{2025}
\doi{10.1093/nsr/nwaf066}
\received{21/10/2024}
\revised{6/12/2024}
\accepted{23/01/2025}

\begin{document}

\dhead{RESEARCH ARTICLE}

\subhead{PHYSICS}

\title{AI-accelerated Discovery of Altermagnetic Materials}

\author{Ze-Feng Gao$^{1,2,\dagger}$}
\author{Shuai Qu$^{2,\dagger}$}
\author{Bocheng Zeng$^{1,\dagger}$}
\author{Yang Liu$^{3}$}
\author{Ji-Rong Wen$^{1}$}
\author{Hao Sun$^{1,*}$}
\author{Peng-Jie Guo$^{2,*}$}
\author{Zhong-Yi Lu$^{2,*}$}



\affil{$^1$Gaoling School of Artificial Intelligence, Renmin University of China, Beijing, China}

\affil{$^2$School of Physics, Renmin University of China, Beijing, China}

\affil{$^3$School of Engineering Science, University of Chinese Academy of Sciences, Beijing, China}
\affil{$^4$A supplementary online appendix is available with this article at the \em{National Science Review} website.}

\authornote{\textbf{Corresponding authors.} Email: haosun@ruc.edu.cn;  guopengjie@ruc.edu.cn; zlu@ruc.edu.cn}
\authornote{Equally contributed to this work.}

\abstract[ABSTRACT]{Altermagnetism, a new magnetic phase, has been theoretically proposed and experimentally verified to be distinct from ferromagnetism and antiferromagnetism. Although altermagnets have been found to possess many exotic physical properties, the limited availability of known altermagnetic materials hinders the study of such properties. Hence, discovering more types of altermagnetic materials with different properties is crucial for a comprehensive understanding of altermagnetism and thus facilitating new applications in the next generation of information technologies, e.g., storage devices and high-sensitivity sensors. 
Since each altermagnetic material has a unique crystal structure, we propose an automated discovery approach empowered by an AI search engine that employs a pre-trained graph neural network to learn the intrinsic features of the material crystal structure, followed by fine-tuning a classifier with limited positive samples to predict the altermagnetism probability of a given material candidate.
Finally, we successfully discovered 50 new altermagnetic materials that cover metals, semiconductors, and insulators confirmed by the first-principles electronic structure calculations. 
The wide range of electronic structural characteristics reveals that various novel physical properties manifest in these newly discovered altermagnetic materials, e.g., anomalous Hall effect, anomalous Kerr effect, and topological property. Noteworthy, we discovered 4 $i$-wave altermagnetic materials for the first time. 
Overall, the AI search engine performs much better than human experts and suggests a set of new altermagnetic materials with unique properties, outlining its potential for accelerated discovery of the materials with targeted properties.}



\maketitle
\section*{Introduction}\label{sec1}
Magnetic materials form a cornerstone of our modern information society. 
Generally, magnetism is categorized into ferromagnetism and antiferromagnetism. 
Recently, based on the spin group formalism~\cite{litvin1974spin}, a new magnetic phase called altermagnetism has been theoretically proposed~\cite{li-prx, liber-prx}, which exhibits numerous novel physical properties~\cite{li-prx,liber-prx,liber-prxm,liber-PRL,Bai-PRL,bose2022tilted,Guo2023,CAHE-2020,Liber-NRM,Zhou-PRB,Feng-NE,MnTe-PRL,guo2023spin,yuan2021prediction,hayami2019momentum,yuan2020giant,chen2023spin}, paving the path way of new avenues in the next generation of information technology. Both altermagnets and conventional antiferromagnets have compensated antiparallel spin sublattices resulting in vanishing net magnetic momentum. The compensated antiparallel spin sublattices are connected by the spin symmetry $\{C_2^{\perp}||I\}$ or $\{C_2^{\perp}||\tau\}$ transformation for conventional antiferromagnets, but by the spin symmetry $\{C_2^{\perp}||R_i\}$ transformation for altermagnets~\cite{li-prx}. 
Here, the symmetry operations at the left and right of the double vertical bar act only on the spin space and lattice space, respectively; the notation $C_2^{\perp}$ represents the $180^{\circ} $ rotation perpendicular to the spin direction; the notations $I, T, R_i$, and $\tau$ denote space inversion, time reversal, rotation/mirror, and fractional translation operations, respectively. Due to the absence of spin symmetry $\{C_2^{\perp}T||IT\}$ or $\{C_2^{\perp}||\tau\}$, altermagnets have spin splitting in electronic bands. Unlike isotropic $\rm{\textbf{k}}$-independent $s$-wave spin splitting in ferromagnets, altermagnets can form anisotropic $\rm{\textbf{k}}$-dependent $d$-wave, $g$-wave, and $i$-wave spin splitting according to different spin group symmetry~\cite{li-prx}. 
Moreover, altermagnets have not only spin-splitting bands deriving from magnetic exchange interaction which is the same as ferromagnets but also unique extraordinary spin-splitting bands deriving from anisotropic electric crystal potential and magnetic exchange interaction~\cite{li-prx}. 
In some altermagnets, the spin splitting can even have magnitudes of $eV$ in parts of the Brillouin zone~(BZ)~\cite{li-prx, liber-prx, liber-prxm}. 
The anisotropic $\rm{\textbf{k}}$-dependent spin splitting can result in a unique spin current by electrical means in $d$-wave altermagnets~\cite{liber-PRL}. Based on the unique spin current, the spin-splitter torque in $d$-wave altermagnets was proposed in theory~\cite{liber-PRL} and confirmed by experiments~\cite{Bai-PRL, bose2022tilted}, which may circumvent limitations of spin-transfer torque~(ferromagnets) and spin-orbit torque~(conventional antiferromagnets or nonmagnetic materials with strong spin-orbit coupling) in magnetic memory devices~\cite{liber-PRL}. Meanwhile, the giant, tunneling magnetoresistance~\cite{liber-prxm} and giant piezomagnetisme~\cite{r15} can also be proposed in altermagnets based on the anisotropic $\rm{\textbf{k}}$-dependent spin splitting. In the relativistic case, the time-reversal symmetry-breaking macroscopic phenomena, including quantum anomalous Hall~\cite{Guo2023}, anomalous Hall~\cite{CAHE-2020, Liber-NRM}, and anomalous Kerr effects~\cite{Zhou-PRB}, have been predicted by theories in altermagnets, moreover, the anomalous Hall effect has been supported by experiments~\cite{Feng-NE, MnTe-PRL}. 

On the other hand, magnetic topological phases and their exotic physical properties have recently attracted intensive experimental and theoretical attention. Very recently, some topological semimetal and insulator phases protected by spin group symmetry have been proposed in theory~\cite{Guo-prl, prx-liu, arxiv-yang, xiao2023spin, jiang2024enumeration, chen2024enumeration, corticelli2022spin}. Considering that the altermagnets are described by spin group symmetry and the symmetry landscape of spin space groups is more plentiful than that of the conventional magnetic space groups, more new magnetic topological phases and their exotic physical properties may be thus proposed theoretically in altermagnets. Nevertheless, altermagnets are hitherto in the early stage of research. Since there are many exotic physical properties that have been discovered and new physical phenomena to be discovered, altermagnets are bound to attract intensive theoretical and experimental attention in the near future. Very recently, based on spin group theory and known magnetic structures, 141 altermagnetic materials have been discovered~\cite{xiao2023spin}. However, known altermagnetic materials are still limited so far. Hence, there is an urgent need to discover more altermagnetic materials for a comprehensive understanding of altermagnetism, thus facilitating new applications in the next generation of information technology.

Conventional discovery methods primarily rely on the known magnetic structures and the corresponding spin space group. Such approaches are applicable only when the magnetic structure information is known a priori, which has a clear limitation if such information is missing. However, there exist over 90,000 magnetic materials documented in the Material Project~\cite{jain2013materials}, among which only 2,138 magnetic structures are known (see the MAGNDATA database~\cite{gallego2016magndata}). The reason why the magnetic structures of only about 2\% of magnetic materials have been determined is that it is indeed a non-trivial task that relies on extremely costly neutron scattering experimentation. Therefore, it is crucial to develop a method that breaks the bottleneck limitation of missing magnetic structure information, enabling the discovery of new altermagnetic materials without any prior knowledge of such information.
On the other hand, the altermagnetic property is closely related to the material crystal structure, which provides a basis for the application of artificial intelligence~(AI) methods to the discovery of altermagnetic materials.  Moreover, the emerging AI technology has found many key applications in the discovery of materials~\cite{wang2023scientific}. For instance, AI was used for predicting organic compound synthesis in organic chemistry~\cite{blakemore2018organic}, planning chemical synthesis pathways~\cite{segler2018planning}, iterative synthesis of small molecules~\cite{lehmann2018towards}, accelerating the discovery of self-assembling peptides~\cite{batra2022machine}, designing eutectic solvent~\cite{luu2023generative}, and analyzing \emph{de novo} protein mechanics and structures~\cite{ni2023generative,ni2024forcegen}.
Recently, deep learning methods have been applied to the prediction of crystal materials with targeted properties~\cite{xie2018crystal,hart2021machine}. 
These methods generally utilize a large amount of crystal structure data to train graph neural network~(GNN) models in an end-to-end manner, without explicit reference to the physical laws underlying these material properties. The trained model could predict key physical properties of crystal materials, such as formation energy and band gap, based on a rich training dataset containing over $10^4$ labeled samples~\cite{xie2018crystal}. However, such methods are not suitable for discovering altermagnetic materials, because of the fact that the known positive samples are limited.

\begin{figure*}[t]
    \centering
    \includegraphics[width=\textwidth]{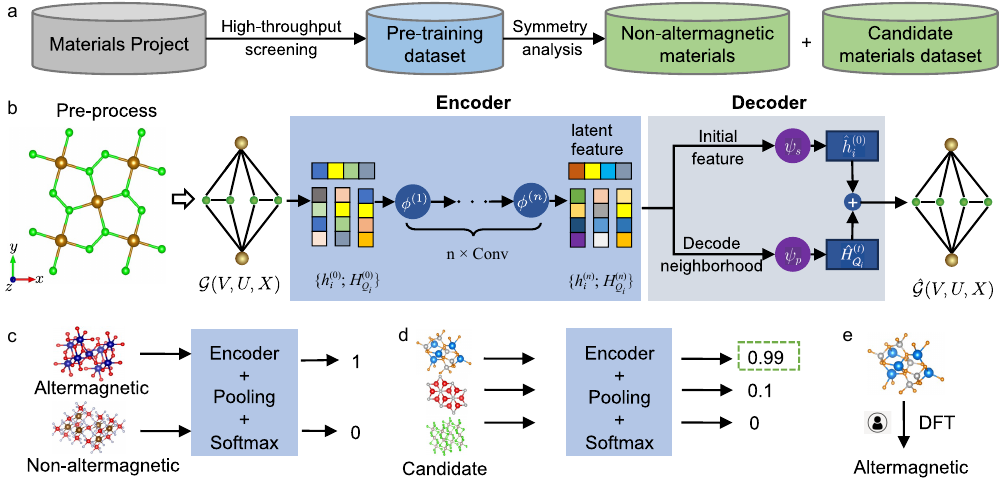}
    \caption{
    {\bf{Workflow of the pre-trained model for searching altermagnetic materials.}}
    {\bf{a}}, Construction of candidate material datasets using high-throughput screening and symmetry analysis (see {Supplementary Appendix Fig. S.1} for details).
    {\bf{b}}, The pre-training autoencoder framework for crystal materials. 
    The input of the model is the crystal structure. Each crystal structure can be represented as a multi-edge graph neural network~(GNN). 
    The encoder is built by the graph convolutional neural network.
    The decoder is built on the Waterstein neighborhood reconstruction.
    {\bf{c}}, The fine-tuning procedure with loading pre-training stage weight matrix.
    {\bf{d}}, The prediction procedure by inputting candidate materials. 
    {\bf{e}}, Validation of the altermagnetic property via the first principle electronic structure calculations.
    }
    \label{fig:main}
\end{figure*}

In this article, we introduce an AI search engine, as shown in Fig. \ref{fig:main}, that combines deep model pre-training and fine-tuning techniques and physics-based approaches~(e.g., symmetry analysis and first-principles electronic structure calculation) to discover new altermagnetic materials under the condition of limited labeled samples. In particular, we pre-train a self-supervised GNN \cite{hu2020strategies} based on optimal transport theory~\cite{ruschendorf1985wasserstein} to learn the intrinsic features of the crystal structure of materials, and refine a downstream classifier with limited positive samples to predict the altermagnetism probability of a given material candidate. First, based on symmetry analysis, we constructed the pre-training dataset~(containing 68,116 materials), fine-tuning dataset~(containing 25,739 materials, namely, 25,591 negative samples plus 148 positive samples), and candidate dataset~(containing 42,377 materials) from the Material Project~\cite{isayev2015materials}. Next, we pre-trained a GNN model (composed of an encoder and a decoder) for crystal materials, based on optimal transport theory. Once the pre-training was done, we fine-tuned the encoder on the fine-tuning dataset to obtain a classifier model. Then, the structured material information from the candidate dataset was input into the classifier model to quantify the probability as an indicator of whether each material is an altermagnetic material. We filtered out the materials with probabilities greater than 0.9 as the candidate altermagnetic materials. Finally, we employed the first-principles electronic structure calculations to estimate the ground magnetic structure of the candidate material to identify altermagnets\footnote{It has been quite common to use the first-principles electronic structure calculations, e.g., density functional theory (DFT), to predict the material property, which has been widely used in the community and proven to possess excellent alignment with experimental results for crystal materials~\cite{deng2018gate,gong2017discovery,klein2018probing,li2021van}.}. Furthermore, the confirmed altermagnetic materials were added to the fine-tuning dataset for an iterative process of fine-tuning and classifier prediction, reinforcing the predictability of the model. The efficacy of this AI search engine has been well demonstrated.

Of 91,649 total candidates, we discovered 50 new altermagnetic materials covering metals, semiconductors, and insulators. The wide range of electronic structural properties implies that various novel physical properties appear in these newly discovered altermagnetic materials, e.g., anomalous Hall effect, anomalous Kerr effect, and topological property, as demonstrated in theoretical analyses. It is also worth noting that we discovered 4 $i$-wave altermagnetic materials for the first time, filling in the gap in the literature. As a result, our proposed AI search engine successfully breaks the bottleneck limitation of existing discovery methods based on symmetry delimited rules, serves as a critical counterpart to such methods, and is applicable to discovering new altermagnetic materials directly from a large set of candidates without any prior knowledge of the magnetic structure information. We conclude that the AI search engine  suggests a set of new altermagnetic materials with unique properties, outlining its potential for accelerated discovery of the materials with targeting properties. We also discuss the pathway of developing pre-trained graph models for the discovery of other types of materials.

\section*{Results}\label{sec-results}
\subsection*{Dataset screening via symmetry analysis}
\label{subsec-symmetry-analysis}
Our goal is to screen altermagnetic materials from the Material Project~\cite{jain2013materials}, which contains 154,718 crystal materials. 
Since this material database includes both magnetic and non-magnetic materials, we first filtered out materials containing magnetic atoms. 
In this work, we considered materials with 3d transition metals or 4f rare earth elements. 
After filtering and de-duplication, we obtained 91,649 potential magnetic materials. 
Due to the complexity of the magnetic properties of materials with multiple magnetic atoms, we further excluded such materials, resulting in 68,116 potential magnetic materials, which constitute the pre-training dataset.

Altermagnetism is characterized by compensated antiparallel spin sublattices connected by the spin symmetry $\{C_2^{\perp}||R_i\}$ transformation but not connected by the spin symmetry $\{C_2^{\perp}||I\}$ or $\{C_2^{\perp}||\tau\}$ transformation.
Since the space groups $P1 (1)$ and $P\bar{1}(2)$ do not have $R_i$ symmetry, all materials with space groups $P1$ and $P\bar{1}$ symmetry are excluded from the pre-training dataset. 
If collinear antiferromagnets have type-IV magnetic space group symmetry, their compensated antiparallel spin sublattices must be connected by the spin symmetry $\{C_2^{\perp}||\tau\}$ transformation in a nonrelativistic case. 
So all collinear antiferromagnetic materials with type-IV magnetic space group symmetry are conventional antiferromagnets but not altermagnets.
Different from antiferromagnetic materials with the type-IV magnetic space group symmetry, the magnetic cell and crystal cell of materials with type-III magnetic space group symmetry are usually the same, which leads to these materials without the spin symmetry $\{C_2^{\perp}||\tau\}$. 
If the magnetic cell of a collinear antiferromagnet is a supercell whereas its spin arrangement breaks the spin symmetry $\{C_2^{\perp}||\tau\}$, then the collinear antiferromagnet may be a supercell altermagnet~\cite{jaeschke2023supercell}. 
Although there exist 4 known supercell altermagnetic materials, we do not consider this situation and exclude them in the positive samples. Since compensated antiparallel spin sublattices in altermagnets require the candidate magnetic materials to have an even number of magnetic atoms in their crystal primitive cell, we first ruled out 18,546 magnetic materials with the odd number of magnetic atoms in the primitive crystal cell from the pre-training dataset. 

Furthermore, magnetic materials with type-III magnetic space group symmetry can be divided into two classes according to space-inversion symmetry. If the crystal structure of a collinear antiferromagnet has no combination of space-inversion and time-reversal symmetry, such a material must be altermagnetic. Otherwise, if the collinear antiferromagnetic material has only a pair of spin antiparallel magnetic atoms in the primitive crystal cell that are not located at invariant space-inversion points, the pair of spin antiparallel magnetic atoms must be connected by the spin symmetry $\{C_2^{\perp}||I\}$. That is to say, the class collinear antiferromagnets are non-altermagnetic materials~(7,045 in total). 
Therefore, based on symmetry analysis, we screened out 25,591 non-altermagnetic materials. These materials, along with the known 148 altermagnetic materials (e.g., as positive samples), constitute the fine-tuning dataset. 
By removing the 25,591 non-altermagnetic materials and positive samples from the pre-training dataset, we obtained the candidate dataset~(42,377 materials). The aforementioned screening process is depicted in Supplementary Appendix Fig. S.1. In the following, we train a neural network to screen and predict altermagnetic materials from the candidate dataset.

\subsection*{Pre-training GNN for materials discovery}

\begin{figure*}[t!]
    \centering
    \includegraphics[width=\textwidth]{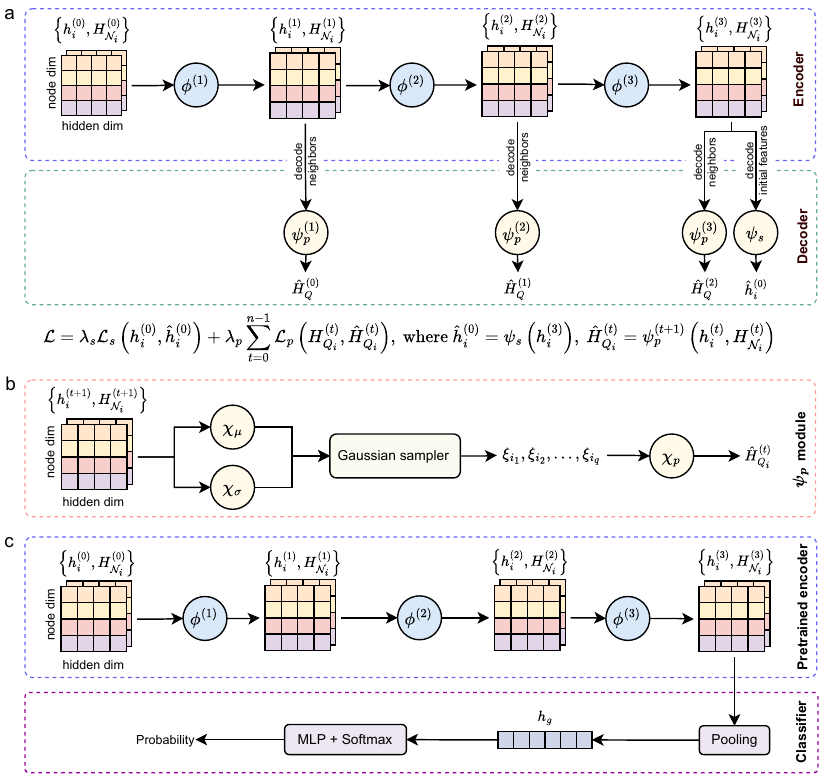}
    \caption{
    \textbf{The network architectures of the auto-encoder and the classifier.} 
    {\bf{a}}, Details of the auto-encoder model. The encoder consists of three graph convolution layers denoted by $\phi^{(1)}, \phi^{(2)}, \phi^{(3)}$, whose input are node features $h_i^{(t)}$ and neighbors features $H_{\mathcal{N}_i}^{(t)}$, where $t=0,1,2$, respectively. The decoder is composed of a decoder module $\psi_s$ for reconstructing initial node features and three decoder modules $\psi_p^{(1)}, \psi_p^{(2)}, \psi_p^{(3)}$ for reconstructing neighborhood set of node features. We minimize the weighted sum of the reconstruction loss functions for both decoder modules. 
    {\bf{b}}, Details of the $\psi_p$ module in the decoder. The $\psi_p$ module includes three MLPs ($\chi_{\mu}, \chi_{\sigma}$ and $\chi_{p}$) and a Gaussian sampler, while the $\psi_s$ module is composed of a single MLP. 
    {\bf{c}}, Details of the classifier model. The node features $h_i^{(0)}$ and the neighborhood set of node features $H_{\mathcal{N}_i}^{(0)}$ are fed into the pre-trained encoder. The output node features ${h}_i^{(3)}$ are then transformed to a latent vector ${h}_g$ by a pooling layer. Finally, another MLP and softmax module is designed to output the probability that quantifies whether the material is altermagnetic.
    }
    \label{fig: appd_model}
\end{figure*}

Although  AI methods have shown great potential for material screening and discovery, there still remain numerous challenges in the field of discovering altermagnetic materials that have not yet been accommodated in existing research practices.
In particular, training a reliable predictive model under the condition of limited labels is intractable, \emph{e.g.}, the number of known altermagnetic materials, as positive samples (training labels), is limited~(only 148 altermagnetic materials \cite{liber-prx,xiao2023spin}). 
We address this challenge by introducing a pre-training and fine-tuning technique, which was first proposed in the natural language processing field~\cite{devlin2018bert}, and subsequently demonstrated with remarkable capabilities for computer vision~\cite{ranftl2021vision} and bioinformatics~\cite{cramer2021alphafold2}. Such a technique pre-trains a self-supervised model first, then refines it for a specific downstream task with limited data meanwhile maintaining a boosted performance. Since each crystal material has a unique pair of structure and property, \emph{e.g.}, the magnetic property like spin pattern is closely related to the crystal structure information, we hypothesize that there is a functional correspondence between the spin pattern and the crystal structure for a given material candidate. Hence, we represent the material crystal structures by mutli-edge graphs and establish a pre-trained neural network model to extract their corresponding latent features. The discovery of altermagnetic materials process is then treated as a downstream task by refining a classifier model based on limited positive samples (\emph{e.g.}, 148 available altermagnetic materials).

As detailed in the symmetry analysis above, we first construct the pre-training dataset and candidate materials dataset based on high-throughput screening and symmetry analysis~(see Fig.~\ref{fig:main}\textbf{a}).
The pre-trained model is based on a GNN that leverages material crystal structure information ~\cite{xie2018crystal}, consisting of a graph convolutional network encoder, and a decoder that reconstructs graph features based on the optimal transport theory~\cite{tang2022graph}. Fig.~\ref{fig:main}{\bf{b}} depicts the schematic of the network, with the detailed architecture shown in Fig.~\ref{fig: appd_model}. 
The process of inputting crystal structures into the model begins with a pre-processing stage, where the crystal structure information is transformed into a graph representation.
Then, we pre-train the model based on the pre-training dataset which contains 68,116 materials, and then fine-tune the pre-trained model based on the fine-tuning dataset~(148 altermagnetic materials plus 25,591 non-altermagnetic materials~\footnote{Note that there are some biases in the negative sampling, but its influence on the predictive performance of our AI model is negligible. This is because the number of negative samples is significantly larger, being 172 times greater than the number of positive samples.})~(see Fig.~\ref{fig:main}{\bf{c}}).
During the fine-tuning, we utilize the pre-trained encoder and employ up-sampling techniques~(duplication and rotation) to balance the number of positive and negative samples for a binary classification task.
Afterward, we can obtain the classifier model, which is then used to screen the altermagnetic materials~(Fig.~\ref{fig:main}{\bf{d}}).
All possible candidate crystal structures~(42,377) are input into the classifier model for prediction. 
The model provides a probability estimate for each sample, and we selected the material with a probability greater than 0.9 as the candidate material.
Next, we utilize the first principle electronic structure calculations~(Fig.~\ref{fig:main}{\bf{e}}) to verify whether the candidates are altermagnetic materials.
Furthermore, once the new altermagnetic materials are verified and confirmed, we add the new one to the fine-tuning dataset and then re-perform the fine-tuning and prediction iteratively. 
Through four rounds of iteration and leveraging information from 148 known altermagnetic materials, we identified 50 new altermagnetic materials. The discussion of the model convergence and the spin patterns distinguish ability is discussed in Supplementary Appendix Note C.
The additional information for the pre-trained model is found in Supplementary Appendix Note A. 

To demonstrate the capability of our pre-trained crystal material model, we fed all the candidate materials in batches into the pre-trained encoder which provides a corresponding latent space vector for each material. We utilized principal component analysis~(PCA) for dimensionality reduction and performed feature visualization and t-SNE visualization on the latent space vectors~(see Supplementary Appendix Fig. S.2). 
The results show that the data in the candidate set have a clear clustering phenomenon after pre-training, which indicates that the pre-training process can group materials containing similar information together.

\begin{table*}[h!]
\centering
\caption{\textbf{Discovered 50 new altermagnetic materials verified by electronic structure calculations.} We further give the nonmagnetic space group, even-parity wave anisotropy, metal~(M) and insulator~(I) conduction type. The altermagnetic $\text{Nd}\text{B}_2\text{C}_2$ is confirmed by previous neutron scattering experiments~\cite{NdBC} and our symmetry analysis. ``NA'' indicates that this material has not been experimentally synthesized. The information of a material whether being a metal or an insulator is confirmed by DFT calculations.
}
\footnotesize
\begin{tabular}{lllllll}
\toprule
{ Number} & { Materials}  & { Space group} & { Anisotropy} & { Conduction} & Material Project ID & References \\ \hline
{1} & {$\text{Nb}_2\text{Fe}\text{B}_2$}    & { $P4$/$mbm~(127)$} & {$g$-wave}     & {M} & mp-1086660 &\cite{Nb2FeB2}         \\
{2} & {$\text{Ta}_2\text{Fe}\text{B}_2$}    & { $P4$/$mbm~(127)$} & {$g$-wave}     & {M} & mp-1095076 &\cite{Ta2FeB2}         \\
{3} &{$\text{Nd}\text{B}_2\text{C}_2$}     & { $P4$/$mbm~(127)$} & {$g$-wave}     & {M} & mp-5765 &\cite{NdBC}          \\
{4} &{$\text{Mg}_2\text{Fe}\text{Ir}_5\text{B}_2$} & { $P4$/$mbm~(127)$} & {$g$-wave}     & {M} & mp-1188243 &\cite{Mg2FeIr5B2}      \\
{5} &{$\text{Mg}_2\text{Mn}\text{Ir}_5\text{B}_2$} & { $P4$/$mbm~(127)$} & {$g$-wave}     & {M} & mp-1189623 &\cite{Mg2FeIr5B2}      \\
{6} &{$\text{Mg}_2\text{Ni}\text{Ir}_5\text{B}_2$} & { $P4$/$mbm~(127)$} & {$g$-wave}     & {M} & mp-1188248 &\cite{Mg2FeIr5B2}      \\
{7} &{$\text{Sc}_2\text{V}\text{Ir}_5\text{B}_2$}  & { $P4$/$mbm~(127)$} & {$g$-wave}     & {M} & mp-20524 &\cite{Sc2VIr5B2}       \\
{8} &{$\text{Sc}_2\text{Mn}\text{Ir}_5\text{B}_2$} & { $P4$/$mbm~(127)$} & {$g$-wave}     & {M} & mp-1208987&\cite{Sc2VIr5B2}       \\
{9} &{$\text{Ca}\text{La}\text{Fe}\text{Ag}\text{O}_6$} & { $Pc~(7)$} & {$d$-wave}     & {M}  & mp-1641528  &NA      \\
{10} &{$\text{Ca}\text{La}\text{Cr}_2\text{O}_6$} & { $Pmn2_1~(31)$}   & {$d$-wave}     & {M}   & mp-1642123    &NA       \\
{11} &{$\text{Ni}\text{F}_3$}     & { $R\bar{3} c~(167)$}     & {$i$-wave}     & {M} & mp-561428 &\cite{FeF3}             \\
{12} &{$\text{Gd}\text{B}_2\text{C}_2$}                 & {$P4/mbm~(127)$}     & {$g$-wave}     & {M} &mp-1080176 &\cite{GdBC} \\
{13} &{$\text{Ho}\text{B}_2\text{C}_2$}                 & {$P4/mbm~(127)$}     & {$g$-wave}     & {M} &mp-20410 &\cite{HoBC} \\
{14} &{$\text{Lu}\text{Cr}\text{O}_3$}                  & {$Pnma~(62)$}     & {$d$-wave}     & {M} &mp-755471 &\cite{LuCrO}             \\
{15} &{$\text{Ta}\text{Co}\text{B}_2$}                  & {$Pnma~(62)$}     & {$d$-wave}    & {M} &mp-1189690 &NA \\
{16} &{$\text{Nd}\text{Ru}\text{O}_3$}                  & {$Pnma~(62)$}     & {$d$-wave}     & {M} &mp-1200843 &\cite{NdRuO}             \\
{17} &{$\text{Fe}\text{H}\text{O}_2$}      & { $Pmn2_1~(31)$}   & {$d$-wave}     & { I} & mp-510670 &\cite{FeHO2}           \\
{18} &{$\text{Na}\text{Fe}\text{O}_2$}     & { $ Pna2_1~(33)$} & {$d$-wave}     & { I} & mp-21060 &\cite{NaFeO2_33}       \\
{19} &{$\text{Na}\text{Fe}\text{O}_2$}     & { $ P4_12_12~(92)$} & {$d$-wave}     & { I} & mp-21880 &\cite{NaFeO2_92}       \\
{20} &{$\text{Mn}\text{O}_2$}     & { $ Pnma~(62)$} & {$d$-wave}     & { I} & mp-19326 &\cite{MnO2_62}        \\
{21} &{$\text{Mn}\text{O}_2$}     & { $ I4/m~(87)$} & {$d$-wave}     & { I} & mp-19395 &\cite{MnO2_87}       \\
{22} &{$\text{Ca}_3\text{Cr}_2\text{O}_7$} & { $Cmc2_1~(36)$}   & {$d$-wave}     & { I}      & mp-1575873 &NA         \\
{23} &{$\text{Zr}\text{Cr}\text{O}_3$}      & { $Pnma~(62)$}   & {$d$-wave}     & { I}      & mp-755055 &NA         \\
{24} &{$\text{Zr}\text{Mn}\text{O}_3$}      & { $R3c~(161)$}   & {$i$-wave}     & { I}       & mp-754513 &NA       \\
{25} &{$\text{V}\text{F}_3$}     & { $R\bar{3} c~(167)$}    & {$i$-wave}     & { I} & mp-559931 &\cite{FeF3}            \\ 
{26} &{$\text{Cr}\text{F}_3$}     & { $R\bar{3} c~(167)$}     & {$i$-wave}     & { I} & mp-560338 &\cite{FeF3}            \\ 
{27} &{$\text{Mn}\text{O}$}     & { $ P6_3mc~(186)$} & {$g$-wave}     & { I} & mp-999539 &\cite{MnO2_186}       \\
{28} &{$\text{Ca}\text{Mn}\text{N}_2$}     & { $P6_3/mmc~(194)$}     & {$g$-wave}     & { I} & mp-1246377 &NA \\
{29} &{$\text{Ba}_2\text{Fe}\text{Ge}_2\text{O}_7$}     & {$P\overline{4}2_1m~(113)$}     & {$g$-wave}     & {~I} &mp-1190820 &\cite{BaFeGeO} \\
{30} &{$\text{Ba}_2\text{Co}\text{Si}_2\text{O}_7$}     & {$P\overline{4}2_1m~(113)$}     & {$g$-wave}     & {~I} &mp-510015 &\cite{BaCoSiO} \\
{31} &{$\text{Sr}_2\text{Co}\text{Ge}_2\text{O}_{7}$}   & {$P\overline{4}2_1m~(113)$}     & {$g$-wave}     & {~I} &mp-1191317 &\cite{SrCoGeO} \\
{32} &{$\text{V}\text{F}_4$}                            & {$P2_1/c~(14)$}     & {$d$-wave}     & {~I} &mp-760030 &NA \\
{33} &{$\text{Ca}_2\text{Co}\text{Te}\text{O}_{6}$}     & {$P2_1/c~(14)$}     & {$d$-wave}     & {~I} &mp-552051 &\cite{CaCoTeO} \\
{34} &{$\text{Ni}\text{F}_2$}                           & {$Pnnm~(58)$}     & {$d$-wave}     & {~I} &mp-556324 &\cite{NiF58} \\
{35} &{$\text{Li}\text{Fe}_2\text{F}_{6}$}              & {$P4_2nm~(102)$}     & {$d$-wave}     & {~I} &mp-557403 &\cite{LiFeF} \\
{36} &{$\text{Fe}\text{H}\text{O}_2$}                   & {$P2_12_12_1~(19)$}     & {$d$-wave}     & {~I} &mp-625251 &NA\\
{37} &{$\text{Ca}\text{Mn}\text{O}_3$}     & {$Pnma~(62)$}     & {$d$-wave}     & {~I} &mp-19201 &\cite{CaMnO}             \\
{38} &{$\text{Ca}\text{V}\text{O}_3$}     & {$Pnma~(62)$}     & {$d$-wave}     & {~I} &mp-22608 &\cite{CaVO}             \\
{39} &{$\text{La}\text{Fe}\text{O}_3$}     & {$Pnma~(62)$}     & {$d$-wave}     & {~I} &mp-22590 &\cite{LaFeO}             \\
{40} &{$\text{La}\text{V}\text{O}_3$}     & {$Pnma~(62)$}     & {$d$-wave}     & {~I} &mp-19350 &\cite{LaVO}             \\
{41} &{$\text{Mn}\text{Se}\text{O}_4$}     & {$Pnma~(62)$}     & {$d$-wave}     & {~I} &mp-817982 &\cite{MnSeO}             \\
{42} &{$\text{Na}\text{Pr}_2\text{Os}\text{O}_6$}     & {$P2_1/c~(14)$}     & {$d$-wave}     & {~I} &mp-20009 &\cite{NaPrOsO}             \\
{43} &{$\text{Na}\text{Pr}_2\text{Ru}\text{O}_6$}     & {$P2_1/c~(14)$}     & {$d$-wave}     & {~I} &mp-542512 &\cite{NaNdRuO}             \\
{44} &{$\text{Nd}\text{Rh}\text{O}_3$}     & {$Pnma~(62)$}     & {$d$-wave}     & {~I} &mp-4582 &\cite{NdRhO}             \\
{45} &{$\text{Pr}\text{Ru}\text{O}_3$}     & {$Pnma~(62)$}     & {$d$-wave}     & {~I} &mp-20186 &\cite{PrRuO}             \\
{46} &{$\text{Sc}\text{V}\text{O}_3$}     & {$Pnma~(62)$}     & {$d$-wave}     & {~I} &mp-756546 &\cite{ScVO}             \\
{47} &{$\text{Sm}\text{Rh}\text{O}_3$}     & {$Pnma~(62)$}     & {$d$-wave}     & {~I} &mp-3317 &\cite{SmRhO}             \\
{48} &{$\text{Ca}\text{La}\text{Cr}\text{Mo}\text{O}_6$}     & {$Pc~(7)$}     & {$d$-wave}     & {~I} &mp-1640189 &NA             \\
{49} &{$\text{La}_2\text{Mn}\text{Rh}\text{O}_6$}     & {$P2_1/c~(14)$}     & {$d$-wave}     & {~I} &mp-1223338 &NA             \\
{50} &{$\text{Li}\text{Fe}\text{F}_4$}     & {$P2_1/c~(14)$}     & {$d$-wave}     & {~I} &mp-755632 &NA             \\

\bottomrule
\end{tabular}
\label{main-tab-results}
\end{table*}

\subsection*{Discovered altermagnetic materials}

Based on the proposed AI search engine, we successfully discovered 50 new altermagnetic materials including 16 metals and 34 insulators~(see Table~\ref{main-tab-results}). The computational results for most of the newly discovered altermagnetic materials~(e.g., the first 23 materials listed in Table~\ref{main-tab-results}) are shown in the Supplementary Appendix Note B and Supplementary Appendix Figs. S.6--S.17.
Moreover, the $d$-wave, $g$-wave, and $i$-wave altermagnets can be found in the predicted 50 altermagnetic materials as shown in Table~\ref{main-tab-results}. In particular, we predicted 4 $i$-wave altermagnetic materials for the first time.

\begin{figure*}[t]
    \centering
    \includegraphics[width=\textwidth]{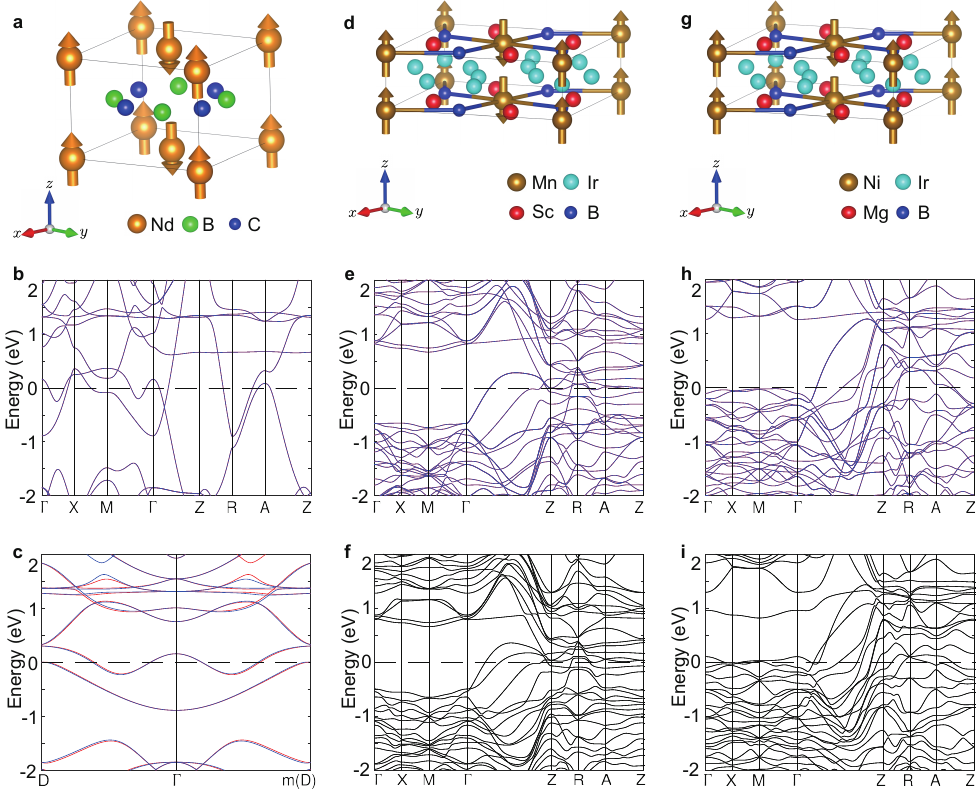}
    \caption{\textbf{The crystal and electronic structure of the altermagnets.}
    {\bf{a}}, The NdB$_2$C$_2$ crystal primitive cell with magnetic structure. 
    {\bf{b}} and {\bf{c}}, The electronic band structure of altermagnetic NdB$_2$C$_2$.  
    The electronic structure is calculated under correlation interaction $\rm{U= 5~eV}$.
    {\bf{d}}, The Sc$_2$MnIr$_5$B$_2$ crystal primitive cell with magnetic structure. 
    {\bf{e}} and {\bf{f}}, The electronic band structure of altermagnetic Sc$_2$MnIr$_5$B$_2$ without and with SOC, respectively. 
    The electronic structure is calculated under correlation interaction $\rm{U= 4~eV}$.
    {\bf{g}}, The Mg$_2$NiIr$_5$B$_2$ crystal primitive cell with magnetic structure. 
    {\bf{h}} and {\bf{i}}, The electronic band structure of altermagnetic Mg$_2$NiIr$_5$B$_2$ without and with SOC, respectively. 
    The electronic structure is calculated under correlation interaction $\rm{U= 6.56~eV}$. 
    The red and blue lines represent spin-up and spin-down energy bands, respectively.
    }
    \label{fig: app-NdBC}
\end{figure*}

The 16 metallic altermagnetic materials can be divided into two classes according to whether the integral of the Berry curvature of the occupied states over the Brillouin zone is zero, which depends on the symmetry of the altermagnetic materials. Since the easy magnetization axes of these materials are in the $x$-$y$ plane, the 8 metallic altermagnetic materials~($\text{Nb}_2\text{Fe}\text{B}_2$, 
$\text{Ta}_2\text{Fe}\text{B}_2$, 
$\text{Nd}\text{B}_2\text{C}_2$,
$\text{Mg}_2\text{Fe}\text{Ir}_5\text{B}_2$, 
$\text{Mg}_2\text{Mn}\text{Ir}_5\text{B}_2$, 
$\text{Mg}_2\text{Ni}\text{Ir}_5\text{B}_2$, 
$\text{Sc}_2\text{V}\text{Ir}_5\text{B}_2$, 
$\text{Sc}_2\text{Mn}\text{Ir}_5\text{B}_2$) have nonzero Berry curvature for the integral of the occupied states over the Brillouin zone according to magnetic point group symmetry, implying that odd-under-time-reversal responses~(e.g., anomalous Hall and Kerr effects) can be realized in these materials. 
Especially, the calculated intrinsic anomalous Hall conductance of altermagnet $\text{Nb}_2\text{Fe}\text{B}_2$ is $-100 \Omega^{-1} cm^{-1}$~\cite{Hou-PRB}, which is the same order of magnitude as those of ferromagnetic metals. 
Since the 3 altermagnetic materials~($\text{Nd}\text{B}_2\text{C}_2$, $\text{Sc}_2\text{Mn}\text{Ir}_5\text{B}_2$, $\text{Mg}_2\text{Ni}\text{Ir}_5\text{B}_2$), whose easy magnetization axes are in the $z$ direction, have zero Berry curvature for the integral of the occupied states over the Brillouin zone, the anomalous Hall effect is not observed. 

Interestingly, the metallic altermagnet $\text{Nd}\text{B}_2\text{C}_2$ has odd-under-time-reversal Dirac fermions protected by the spin symmetry $\{E||C_{4z}\}$ and $\{C_2^{\perp}||M_{x}(\frac{1}{2},\frac{1}{2})\}$ (see Fig.~\ref{fig: app-NdBC}{\bf{b}}), but $\text{Sc}_2\text{Mn}\text{Ir}_5\text{B}_2$ and $\text{Mg}_2\text{Ni}\text{Ir}_5\text{B}_2$ have odd-under-time-reversal sixfold degenerate fermions~(see   Fig.~\ref{fig: app-NdBC}{\bf{e}} and {\bf{h}}) on the $\rm{\Gamma-Z}$ around the Fermi level, which is protected by the spin point group symmetry. 
When considering spin-orbit coupling (SOC), the 3 metallic altermagnets $\text{Nd}\text{B}_2\text{C}_2$, $\text{Sc}_2\text{Mn}\text{Ir}_5\text{B}_2$, and $\text{Mg}_2\text{Ni}\text{Ir}_5\text{B}_2$ have $D_{4h}$ point group symmetry which must be broken in ferromagnets, and the $C_{4v}$ double point group symmetry protects the odd-under-time-reversal Dirac fermions of the metallic altermagnets $\text{Sc}_2\text{Mn}\text{Ir}_5\text{B}_2$ and $\text{Mg}_2\text{Ni}\text{Ir}_5\text{B}_2$ on the $\rm{\Gamma-Z}$~(see   Fig.~\ref{fig: app-NdBC}{\bf{f}} and {\bf{i}}). Moreover, the pair odd-under-time-reversal Dirac points in $\text{Mg}_2\text{Ni}\text{Ir}_5\text{B}_2$ are very close to the Fermi level, which is an advantage for investigating its novel physical properties in experiments.  

\begin{figure*}[t]
    \centering
    \includegraphics[width=\textwidth]{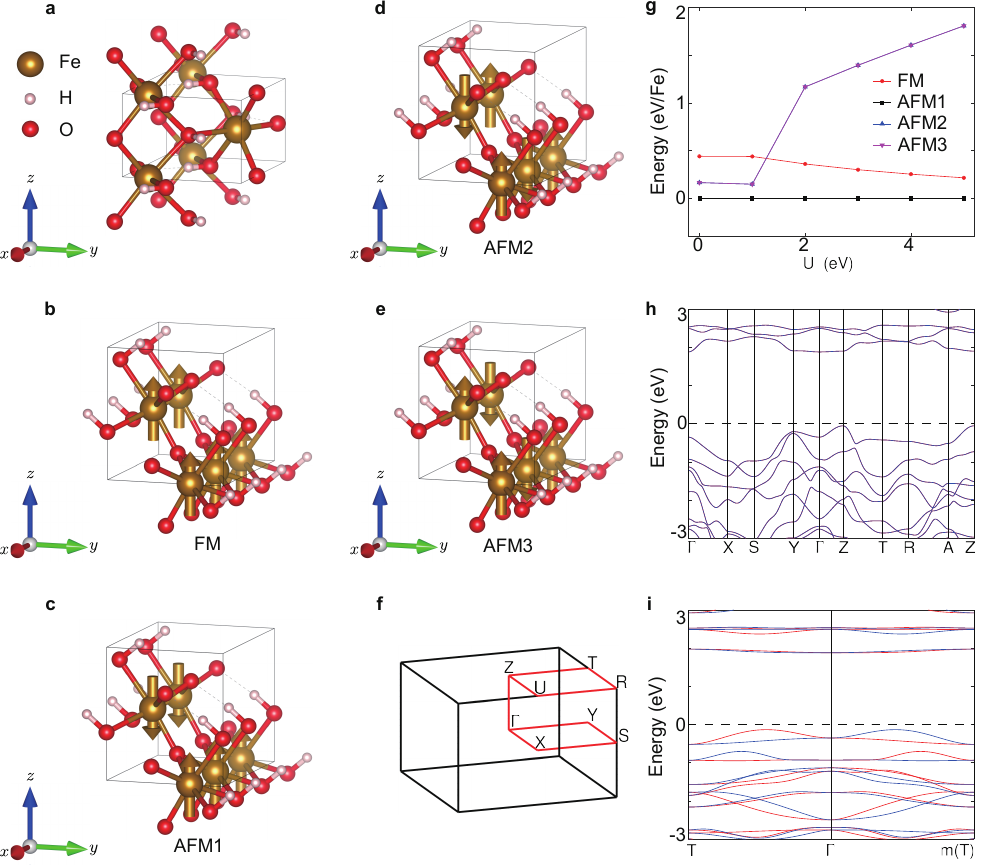}
    \caption{
    \textbf{The crystal and electronic structure of the altermagnet FeHO$_2$~{(31)}.}
    {\bf{a}}, The crystal primitive cell of the altermagnetic FeHO$_2$~{(31)}. 
    {\bf{b}}--{\bf{e}}, Four significant magnetic structure of FeHO$_2$~{(31)}.
    The arrows represent the magnetic moments of Fe. 
    {\bf{f}}, The Brillouin zone (BZ) with high-symmetry points and lines of altermagnetic FeHO$_2$~{(31)}.
    {\bf{g}}, The relative energy of four significant magnetic states with the variation of correlation interaction U. 
    {\bf{h}} and {\bf{i}}, The electronic band structure of FeHO$_2$~{(31)} without SOC. 
    The red and blue lines represent the spin-up and spin-down energy bands, respectively.
    The electronic structure is calculated under correlation interaction $\rm{U= 4~eV}$. }
    \label{fig: FeHO2}
\end{figure*}

On the other hand, ferromagnetic semiconductors that have spintronic and transistor functionalities could be applied to the next generation of electronic devices. However, the ferromagnets usually are metals with a high Curie temperature and hold no brief for insulators with a high Curie temperature. Altermagnets with compensated antiparallel sublattices not only are in favor of insulators with a high Neel temperature but also have spintronic functionality~\cite{liber-prx}. Thus, altermagnets open a new pathway to bypass the difficulties of ferromagnets. 
Here, we employed the LDA+U method~\cite{PhysRevB.57.1505} to predict 34 altermagnetic semiconductors~(see Supplementary Appendix Table S.2).
Furthermore, the altermagnetic $\text{Fe}\text{H}\text{O}_2$ may be a spin-triplet excitonic phase. 
From  Fig.~\ref{fig: FeHO2}{\bf{i}}, we observe that there is large spin splitting of 0.39 eV on the $\rm{T-\Gamma-T}$ directions and the spins of valence and conduction band are opposite, which may result in the spin-triplet excitonic phase~\cite{duan-prl}. Moreover, the energy of altermagnetic state~(AFM1) is much lower than that of the other three magnetic states~(see  Fig.~\ref{fig: FeHO2}{\bf{g}}), indicating that $\text{Fe}\text{H}\text{O}_2$ may have a Neel temperature above the room temperature. Thus, the altermagnetic $\text{Fe}\text{H}\text{O}_2$ is a very interesting material that, we believe, will attract both theoretical and experimental interests. In addition, although $\text{Nd}\text{Ru}\text{O}_3$ is a altermagnetic semimetal, it has bandgap along the high-symmetry directions with spin splitting, and its valence and conduction bands have opposite spins ({Supplementary Appendix Fig. S.5}{\bf{e}} and {\bf{f}}). Thus, $\text{Nd}\text{Ru}\text{O}_3$ may be a Bardeen Cooper Schrieffer (BCS) type triplet exciton insulator \cite{zhang2024spontaneous}. In the following, we present in detail two altermagnetic materials which are metal and semiconductor, respectively.

\begin{figure*}[h!]
    \centering
    \includegraphics[width=\textwidth]{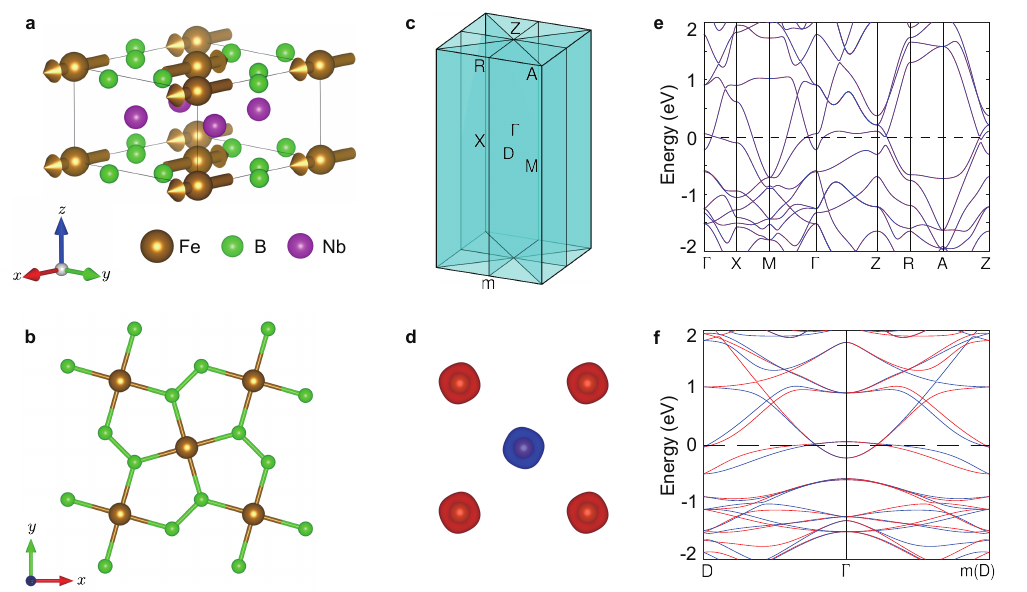}
    \caption{
    {\textbf{The crystal and electronic structure of the altermagnetic $\text{Nb}_2\text{Fe}\text{B}_2$.} }
    {\bf{a}}, The side view of altermagnetic $\text{Nb}_2\text{Fe}\text{B}_2$.
    {\bf{b}}, The top view of altermagnetic $\text{Nb}_2\text{Fe}\text{B}_2$. 
    {\bf{c}}, The Brillouin zone~(BZ) with high-symmetry points of altermagnetic $\text{Nb}_2\text{Fe}\text{B}_2$. The cyan plane represents the nodal surface of BZ. The m represents the mirror symmetry $M_y$. 
    {\bf{d}}, The anisotropic spin-charge density deriving from an anisotropic crystal field. 
    {\bf{e}}, The electronic band structure along high-symmetry directions without spin-orbit coupling~(SOC). 
    {\bf{f}}, The electronic band structure along non-high-symmetry directions without SOC.  
    The red and blue lines represent the spin-up and spin-down energy bands, respectively.
    The electronic structure is calculated under correlation interaction $\rm{U= 4.82~eV}$.
    }
    \label{fig: Nb2FeB2}
\end{figure*}

$\text{Nb}_2\text{Fe}\text{B}_2$ has space group $P4$/$mbm~(127)$ symmetry, and the corresponding elementary symmetry operations are $C_{4z}$,$ C_{2x} (\frac{1}{2}, \frac{1}{2})$ and $I$, which yield the point group $D_{4h}$. 
The crystal structure of $\text{Nb}_2\text{Fe}\text{B}_2$ is composed of $Fe-B$ atoms layer and $Nb$ atoms layer as shown in Fig.~\ref{fig: Nb2FeB2}{\bf{a}}. Moreover, the two $Fe$ atoms in the primitive cell are surrounded by two $B$ atomic quadrilaterals with different orientations, respectively~(Fig.~\ref{fig: Nb2FeB2}{\bf{b}}). 
Very recentlly, $\text{Nb}_2\text{Fe}\text{B}_2$ has been predicted to be a Neel antiferromagnet, which is shown in Fig.~\ref{fig: Nb2FeB2}{\bf{a}}. Due to the anisotropic $Fe-B$ quadrilateral, the spin-charge density of $Fe$ atoms is anisotropic (see Fig.~\ref{fig: Nb2FeB2}{\bf{d}}). Thus, compensated antiparallel spins are not connected by the spin symmetry $\{C_2^{\perp}||I\}$ or $\{C_2^{\perp}||\tau\}$ but are connected by the spin symmetry$\{C_2^{\perp}||C_{2x}(\frac{1}{2},\frac{1}{2})\}$; that is to say, $\text{Nb}_2\text{Fe}\text{B}_2$ is an altermagnetic material. 
The spin symmetry $\{C_2^{\perp}||M_{x}(\frac{1}{2},\frac{1}{2})\}$ protects the spin degeneracy in electronic bands on the $k_x= 0$ and $\pi$ planes, considering the spin symmetries $\{E||C_{4z}\}$, the altermagnetic $\text{Nb}_2\text{Fe}\text{B}_2$ has six node surfaces in the Brillouin zone (see Fig.~\ref{fig: Nb2FeB2}{\bf{c}}). 
Thus, $\text{Nb}_2\text{Fe}\text{B}_2$ is a $g$-wave altermagnet described by the non-trivial spin Laue group $P^14/^1m^2m^2m$. Fig.~\ref{fig: Nb2FeB2}{\bf{e}} shows that the electronic bands of the altermagnetic $\text{Nb}_2\text{Fe}\text{B}_2$ are spin degenerate along the high-symmetry directions, which is consistent with our symmetry analysis. 
As can be seen from the Fig.~\ref{fig: Nb2FeB2}{\bf{f}}, all the bands are spin-splitting and spin antisymmetric in the non-high-symmetry $D-\Gamma-m(D)$ direction, which reflects the characteristics of $g$-wave spin polarization. On the other hand, the valence bands and the conduction bands have multiple crossing points on the high-symmetry and non-high-symmetry directions, such as the $\Gamma-X$ and $\Gamma-D$ directions, indicating that the altermagnet $\text{Nb}_2\text{Fe}\text{B}_2$ is a topologically nontrivial metal. When considering SOC, the easy magnetization axis of altermagnet $\text{Nb}_2\text{Fe}\text{B}_2$ is along the $x$ direction. 
Accordingly, the altermagnet $\text{Nb}_2\text{Fe}\text{B}_2$ has $C_{2z}T, C_{2x}(\frac{1}{2}, \frac{1}{2})T, C_{2y}(\frac{1}{2}, \frac{1}{2})T, I, M_zT, M_x(\frac{1}{2}$, $\frac{1}{2})T$, $M_y$ point symmetries, which make the anomalous Hall conductance both $\sigma_{xy}$ and $\sigma_{yz}$ zero, but $\sigma_{xz}$ non-zero, which has been predicted by our previous theoretical study~\cite{Hou-PRB}. 
Likewise, the anomalous Kerr effects can also be realized in the altermagnet $\text{Nb}_2\text{Fe}\text{B}_2$. 

\begin{figure*}[t!]
    \centering
    \includegraphics[width=\textwidth]{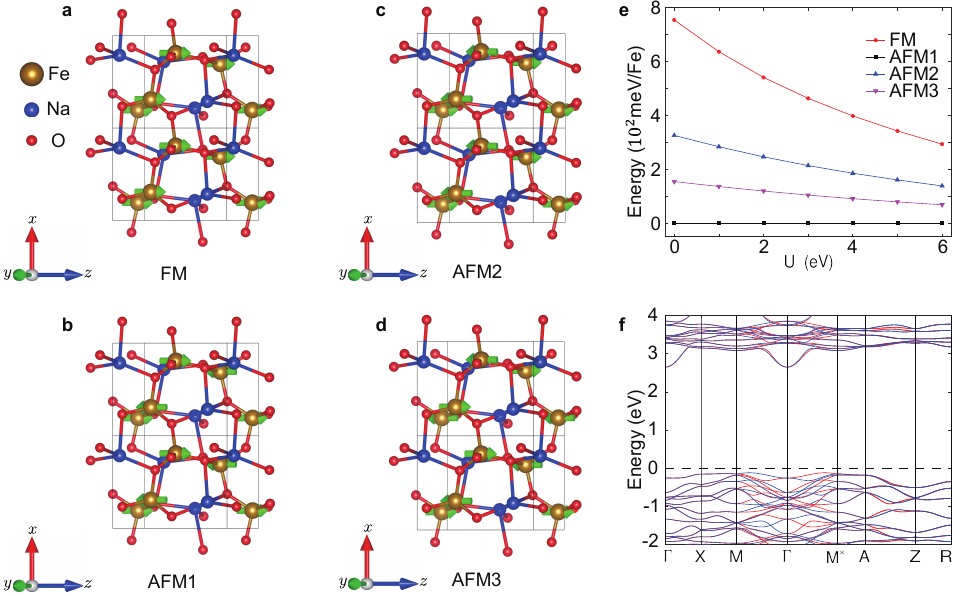}
    \caption{
    {\textbf{The crystal and electronic structure of the altermagnetic $\text{Na}\text{Fe}\text{O}_2~(92)$.} }
    {\bf{a}}, The crystal structures of the $\text{Na}\text{Fe}\text{O}_2~(92)$ with FM structures.
    {\bf{b}}, The crystal structures of the $\text{Na}\text{Fe}\text{O}_2~(92)$ with AFM1 structures.
    {\bf{c}}, The crystal structures of the $\text{Na}\text{Fe}\text{O}_2~(92)$ with AFM2 structures.
    {\bf{d}}, The crystal structures of the $\text{Na}\text{Fe}\text{O}_2~(92)$ with AFM3 structures.
    {\bf{e}}, Relative energy of different magnetic states with the variation of correlation interaction $U$. 
    {\bf{f}}, The electronic band structure of $\text{Na}\text{Fe}\text{O}_2~(92)$ along high-symmetry directions without SOC. 
    The red and blue lines represent the spin-up and spin-down energy bands, respectively. 
    The electronic structure is calculated under correlation interaction $\rm{U= 4~eV}$.}
    \label{fig: NaFeO}
\end{figure*}

The other altermagnetic material that we would like to mention is $\text{Na}\text{Fe}\text{O}_2$. The crystal structure of $ \text{Na}\text{Fe}\text{O}_2$ is shown in Fig.~\ref{fig: NaFeO}{\bf{a}}-{\bf{d}} with space group $P/4_12_12~(92)$ symmetry. The corresponding elementary symmetry operations are $ C_{4z} (\frac{1}{2}, \frac{1}{2}, \frac{1}{4})$ and $ C_{2x} (\frac{1}{2}, \frac{1}{2}, \frac{3}{4})$, which yield the point group $ D_4$. 
Since the $d$ orbitals of $ Fe$ are half occupied and the angle between $ Fe-O-Fe$ is 136 degrees in $\text{Na}\text{Fe}\text{O}_2$, the superexchange interactions result in the nearest neighbor $Fe$ ions with opposite magnetic moments and the next neighbor $Fe$ ions with the same magnetic moments. 
Hence, the magnetic ground state of $\text{Na}\text{Fe}\text{O}_2$ will be the AFM1 (see Fig.~\ref{fig: NaFeO}{\bf{b}}). 
To verify our theoretical analysis, we consider four different magnetic structures, which are shown in Fig.~\ref{fig: NaFeO}{\bf{a}}-{\bf{d}}. 
It can be seen that the magnetic structure AFM1 is always in the ground state of $\text{Na}\text{Fe}\text{O}_2$ under different correlation interactions $U$~(see Fig.~\ref{fig: NaFeO}{\bf{e}}). 
Moreover, the energy of the AFM1 state is much lower than that of the other three magnetic states~(Fig.~\ref{fig: NaFeO}{\bf{e}}), implying that $\text{Na}\text{Fe}\text{O}_2$ may have a Neel temperature above the room temperature. It is shown in Fig.~\ref{fig: NaFeO}{\bf{b}} that the magnetic and crystal primitive cells of $\text{Na}\text{Fe}\text{O}_2$ are the same, which break $\{C_2^{\perp}||\tau\}$ spin symmetry. 
Thus, $\text{Na}\text{Fe}\text{O}_2$ is an altermagnetic material due to the lack of space-inversion symmetry. 

We also calculated the electronic band structure along the high-symmetry directions. Fig.~\ref{fig: NaFeO}{\bf{f}} shows that the altermagnet $ \text{Na}\text{Fe}\text{O}_2$ is a semiconductor with a band gap of 2.75 eV. The spin-degenerate bands on the $\Gamma-{\rm{X,~M}}^{*}-{\rm{A}}$ and $\rm{Z-R}$ directions (the $\rm{X-M}$ direction) are protected by the spin symmetry $\{C_2^{\perp}||C_{2y} (\frac{1}{2}, \frac{1}{2}, \frac{1}{4})\}$ ( $\{C_2^{\perp}||C_{2y} (\frac{1}{2}, \frac{1}{2}, \frac{3}{4})\}$)  (see Fig.~\ref{fig: NaFeO}{\bf{f}}). 
In fact, the spin symmetry $\{T||C_{2y}T (\frac{1}{2}, \frac{1}{2}, \frac{1}{4})\}$~( $\{T||C_{2x}T (\frac{1}{2}, \frac{1}{2}, \frac{3}{4})\}$) can protect spin degeneracy of bands on the $k_y =0$ and $\pi$ (the $k_x = 0$ and $\pi$) planes. 
That is to say, the altermagnet $\text{Na}\text{Fe}\text{O}_2$ has four nodal surfaces in the Brillouin Zone. Thus, $\text{Na}\text{Fe}\text{O}_2$ is a $d$-wave altermagnet which is reflected by the spin-splitting bands on the $\rm{M}-\Gamma-\rm{M}^*$ directions. Considering the $d$-wave altermagnets allowing unique spin current by electrical means \cite{liber-PRL}, the altermagnet $\text{Na}\text{Fe}\text{O}_2$ may have both spintronic and transistor functionalities at the room temperature.  

\section*{Discussion}\label{sec12}

AI approaches have shown ground-breaking capabilities in the discovery of materials in a large search space. An intractable challenge faced by AI lies in the shortage of sufficient labels or positive samples, \emph{e.g.}, in the case of the discovery of altermagnetic materials. We herein introduced an AI search engine that combines pre-trained crystal models~(GNN pre-training and optimal transport theory) and physics-based methods~(symmetry analysis and first-principles electronic structure calculations) to discover new altermagnetic materials with specific properties under minimal labeled sample conditions. Among 91,649 possible candidates, we identified 50 new altermagnetic materials covering metal, semiconductor, and insulator.
Meanwhile, the proposed AI search engine also has the few-shot learning ability. For example, it is capable of predicting 25 altermagnetic materials only based on 14 positive samples (see {Supplementary Appendix Note C}).
We observed various novel physical properties in these newly discovered altermagnetic materials, \emph{e.g.}, anomalous Hall effect, anomalous Kerr effect, and topological property.  It is noted that 4 out of these 50 altermagnetic materials are $i$-wave types, which are discovered for the first time filling a gap in the literature. We demonstrate that the AI search engine is capable of uncovering a set of altermagnetic materials with unique properties, highlighting its potential for accelerated discovery of the materials with targeting properties. 

There still remain some potential limitations associated with the AI search engine. Firstly, we have to admit that the issue of imbalance between positive and negative samples during the fine-tuning stage exists, primarily due to the scarcity of known positive samples. We have also discussed the potential intrinsic error and computational cost associated with this AI search engine~(see {Supplementary Appendix Note C}). Utilizing the translational and rotational symmetries of crystals to augment positive sample data may help address this challenge, which will be demonstrated in our future work. Secondly, if the materials exhibit distinct magnetic phases at varying temperatures, such as Mn$_5$Si$_3$, whose low-temperature spin pattern is reported to be non-collinear and high-temperature spine pattern as antiferromagnetic ~\cite{reichlova2020macroscopic}, our current model is unable to predict their magnetic properties. In such a case, if the material with temperature-driven magnetic transition is used as the positive sample to train the AI model, the prediction might involve possible bias. We will improve our classifier model by considering temperature as a conditional input parameter to enhance its capability in screening and predicting materials with temperature-driven properties in the future. Thirdly, in the calculations of magnetic ground states, we focus only on collinear magnetic structures and do not consider non-collinear ones. This is because most of the materials we calculate do not exhibit geometric frustration. Even for a few materials like CoF$_3$, which exhibit triangular geometric frustration, previous studies did not consider non-collinear magnetic structures~\cite{FeF3}. Another limitation is that we have not yet found ideal altermagnetic topological insulators and altermagnetic topological semimetals (such as odd-under-time-reversal Dirac, and sixfold semimetals). Employing the decoder based on the pre-trained model to generate potential altermagnetic materials holds promise in overcoming this challenge. Furthermore, adopting a multimodal pre-training approach offers the potential to further enhance the accuracy of model predictions. The current pre-training only considers the single modality of crystal structure information. Leveraging information from other modalities~(such as textual descriptions of crystal structures) may enhance the performance of the pre-trained model. These methods will be further explored in our future research endeavors.

In addition, an alternative to alter the proposed AI model is to replace the classifier with a regressor to predict the magnetic structure of a given material candidate, where such a regressor can be fine-tuned based on 2,138 known magnetic structures in the MAGNDATA database~\cite{gallego2016magndata}. Afterward, symmetry analysis can be employed to identify the altermagnetism. However, since there are infinite types of magnetic structures, accurately predicting the exact type of magnetic structure for a given material remains challenging. Therefore, we trust an end-to-end classifier in our proposed model, to directly judge whether a crystal material is altermagnetic, is preferred especially under the condition of very limited positive samples. 

It is well known in the community that the altermagnetism of 98\% of candidates in the Material Project ($>$90,000 materials), whose magnetic structure information is unknown, has not yet been confirmed and remains a substantially challenging task. The brute-force approach leveraging our experts’ knowledge relies on trial and error by chance, having an extremely low probability of correctly discovering and confirming altermagnetic materials from the large database. However, our AI model narrows down the search space, predicts a list of highly possible candidate altermagnets, and lifts the discovery accuracy to a notable margin of about 31\% (50/161), which greatly accelerates our discovery of new altermagnetic materials.~\footnote{If we calculate the magnetic ground state of all materials, and then determine whether these materials are altermagnetic by spin symmetry, we probably can predict more altermagnetic materials. However, this cost is extremely huge. Given the scale and complexity of this task, it is unlikely to use DFT to complete the above brute-force computation for all candidate materials.}
The success of this engine lies not only in its predictive capabilities but also in its ability to leverage extensive crystal structure data and deep learning techniques, allowing for pre-training without explicit reference to underlying physical laws, to reveal complex correlations and patterns in new materials.
Although 161 altermagnetic materials have been confirmed by the symmetry analysis approach~\cite{xiao2023spin}, the urgency of discovering a variety of new altermagnetic materials with different properties still remains. 
Based on these existing altermagnetic materials confirmed by the spin space group, our AI search engine could predict many more altermagnetic materials, among which we expect to find a variety of altermagnets beyond the MAGNDATA database. For example, over 300 additional candidate materials (unconfirmed yet by DFT calculations) were predicted by AI search engine and listed in our GitHub repository\footnote{\url{https://github.com/zfgao66/MatAltMag}} for open research. Nevertheless, obtaining the magnetic ground state without experimental validation remains a challenging problem. The magnetic state of a material may be collinear or non-collinear, and the magnetic cell may be a supercell of the crystal primitive cell, which, in principle, lead to an infinite number of possibilities for the magnetic structure. Hence, experimental efforts will help further validate our discovery.

The proposed engine might be also applicable to other types of materials whose properties are strongly correlated to their crystal structures, such as the Bardeen-Cooper-Schrieffer (BCS) superconductive materials~\cite{pickett2023colloquium}, ferromagnetic semi-conductor~\cite{jungwirth2006theory}, and high-temperature superconductive materials~\cite{varma2020colloquium}, among others.
We will demonstrate the potential of our proposed model for discovering other materials in our future study.
We envision this effort may present new opportunities in the field of material discovery across different disciplines.

\section*{Methods}

We herein introduce the model details and implementation specifics.

\subsection*{Model details}

\paragraph{Architecture overview}
The concept of pre-training a large deep learning model and subsequently applying it to perform downstream tasks originally originated in the field of natural language processing~(NLP). 
Large-scale NLP models, such as GPT~\cite{radford2019language}, and their derivatives, employ transformers as text encoders. 
These encoders transform input texts into embeddings and establish pre-training objectives based on these embeddings, including generative loss and masked language modeling loss.
The pre-training process is typically unsupervised, based on large-scale unlabeled samples. 
In contrast to traditional end-to-end neural network models, pre-trained models can achieve excellent performance even with limited labeled positive samples. 
We thus consider utilizing the pre-training technique to fully leverage the information from existing crystal materials databases and treat the discovery of altermagnetic materials as a downstream task.

The objective of our proposed pre-training model for crystal materials is to learn the information embedded within crystal structures. 
To enhance the learning capacity of the pre-training model, we proposed a graph auto-encoder architecture (see Fig. \ref{fig:main} and   Fig. \ref{fig: appd_model}). 
The encoder consists of $n$ layers of graph convolution to learn crystal embeddings, while the decoder employs the Wasserstein distance based on the optimal transport theory~\cite{ruschendorf1985wasserstein} for the reconstruction of the input crystal structures. 
Specifically, the encoder aims to encode the graphical representation of crystal materials into a high-dimensional matrix, while the goal of the decoder is to decode this one back into the graphical representation of crystal materials. 
Through extensive training with unlabeled data, the model effectively converges~(as depicted in the pre-training loss history as shown in~{Supplementary Appendix Fig. S.3}). 
We believe that the pre-trained encoder can effectively project the crystal structures into crystal embeddings. 
Leveraging the encoder of the pre-trained model, we built the classifier model by incorporating a pooling layer and a softmax function. 
Subsequently, we trained the classifier model using the fine-tuning dataset. This trained model is then employed to screen the candidate materials, offering the probability of whether the target material is altermagnetic.
The hyperparameters of the model were obtained by grid search, as listed in~{Supplementary Appendix Table S.1}.
In summary, our model comprises four main components: crystal data pre-processing, an encoder constructed using graph convolutional neural networks, a decoder built based on optimal transport theory, and the construction of a classifier model.
We elaborate on each of these components one by one as follows.

\paragraph{Crystal data pre-processing}
The data pre-processing procedure aims to bridge the crystal structure and the crystal graph representation \cite{xie2018crystal}.
The input of the model is a crystal structure file~(.cif) that contains three primitive translation vectors of the primitive unit cell and the positional information of each atom.
It satisfies the organization invariance for atomic indexing and the size invariance for unit cell selection. We define the graph representation $\mathcal{G}(V, U, X)$ to describe the crystal structure information, where $V$ denotes the set of nodes, $U$ the set of edges, and $X$ the set of features.
First, we represent atoms as nodes~$v_i$ in a crystal graph representation, where $i=1, ..., |V|$. Since periodic boundary conditions are taken into consideration, equivalent nodes are merged to obtain irreducible nodes.
Then, for each node $v_i$, we consider the neighborhood nodes $v_j$, where $j=,1,...,|\mathcal{N}_i|$ and $\mathcal{N}_i$ the set of neighborhood nodes for $v_i$. The $k$ connections between nodes $v_i$ and $v_j$ are denoted as the edge $u_{(i,j)_k}$ in the graph.
Next, the initial node features $\big\{h_{i}^{(0)}\big\}_{i=1}^{|V|}$ are given through one-hot encoding based on the sequence of atoms in the crystal structure.
We use $H_{\mathcal{N}_i}^{(0)}$ to denote the neighbor node features of node $v_i$.
Here, $v_i$ denotes the $i$-th node.
Each edge $u_{(i,j)_k}\in U$ is represented by a feature vector $u_{(i,j)_k}$ that corresponds to the $k$-th bond linking node $v_i$ and node $v_j$.
A feature vector $h_i\in X$ encoding the attribute of the atom corresponding to node $v_i\in V$ is used to represent each node $v_i$. An example for determining the atom connectivity is illustrated in~{Supplementary Appendix Fig. S.4}. 

\paragraph{Crystal graph convolutional encoder}
The encoder is used to represent the input crystal structure information as a high-dimensional matrix~(Fig.~\ref{fig: appd_model}$\bf{c}$), which contains $n$ convolutional layers.
The $t$-th convolutional layer updates the node feature vector $h_i^{(t)}$ via convolution function $h_i^{(t+1)}={\rm{Conv}}\big(h_i^{(t)},h_j^{(t)},u_{(i,j)_k}\big)$.
We denote the graph convolution function with $g$, which iteratively updates the overall feature vector $h_i$, whose output is the input for the next step. 
The node index in feature vector $h_i$ and length of $h_i$ are invariant for every step, We construct the first concatenate neighbor vector as
$z^{(t)}_{(i,j)_k}=h^{(t)}_i\oplus h^{(t)}_j \oplus u_{(i,j)_k}$ in step $t$, and then perform the convolution operation to update the feature as follows:
\begin{equation}
    h_i^{(t+1)}=h_i^{(t)}+ 
        \sum_{v_j\in \mathcal{N}_i,v_m\in \mathcal{M}_i,k}\sigma\left(A\right)\odot  g\left(B\right),
    \label{eq: encoder}
\end{equation}
where $A$ denotes $z^{(t)}_{(i,j)_k} W_f^{(t)}+b_f^{(t)}$ and $B$ denotes the $z^{(t)}_{(i,j)_k} W_s^{(t)}+h_{i,m}^{(t)}W_m^{(t)}+b_s^{(t)}$. The $\odot$ denotes the element-wise multiplication, $\mathcal{M}_i$ the magnetic atoms corresponding to node $v_i$, and $\sigma$ the sigmoid activation function.
Since the magnetic atoms are important for the material to exhibit altermagetic properties, we added the weight term $W_m^{(t)}$ for the magnetic atoms.
The weight function $W_c^{(t)}, W_s^{(t)}, W_m^{(t)}$ are the convolution weight matrix, self-weight matrix, and magnetic atom weight matrix of the $t$-th layer, respectively. 
In Eq.~\eqref{eq: encoder}, we incorporate the residual term $h_i^{(t)}$ to enhance the training of the neural network.

\paragraph{Neighborhood Wasserstein reconstruction decoder}
The decoder (denoted by $\psi$) is utilized to restore the input graph representation of a crystal from the crystal embeddings, which mainly consists of two parts (see   Fig.~\ref{fig: appd_model}\textbf{a}), one for node feature reconstruction (denoted by $\psi_s$) and the other for adjacent node feature reconstruction (denoted by $\psi_p$), namely, $\psi=(\psi_p+\psi_s)$. Here, $\psi_s={\rm{MLP}}_s\big(h_{i}^{(t)}\big)$ is used to reconstruct the node features, where MLP indicates a multilayer perception. The architecture of the decoder block, as shown in   Fig.~\ref{fig: appd_model}\textbf{b}, follows the design in~\cite{tang2022graph}.

In particular, we adopt the $n$-hop neighboring Wasserstein decoder for graph feature reconstruction.
We can obtain $\big\{h_i^{(0)},H_{\mathcal{N}_i}^{(0)}\big\}$ from the pre-processing procedure.
For each node $v_i \in V$, we update the node representation $h_i^{(t+1)}$ via the GNN layer in the encoder, which gathers information from $h_i^{(t)}$ and its neighbor representations $H_{\mathcal{N}_i}^{(t)}$, namely, $h_i^{(t+1)} = \phi^{(t)} \big(h_i^{(t)},H_{\mathcal{N}_i}^{(t)}\big)$. Note that the neighborhood set of node features $H_{\mathcal{N}_i}^{(t)}$ can be directly assembled based on the node adjacency.
Consequently, we solve the following optimization problem to train the network:
\begin{equation}
    \arg\min_{\phi,\psi}\sum_{v_i \in V}\mathcal{L}
    \left(h_{i}^{(t)}, H_{\mathcal{N}_i}^{(t)}, \psi\left(h_i^{(t+1)}, H_{\mathcal{N}_i}^{(t+1)}\right)\right),
\end{equation}
where $\mathcal{L}(\cdot,\cdot)$ denotes the reconstruction loss over $0\leq t< n$.
The loss function $\mathcal{L}$ can be decomposed into two distinct elements, each gauging the reconstruction of self and neighborhood node features, respectively, written as

\small\begin{equation}
\mathcal{L}=\lambda_s\mathcal{L}_{s}\left(h_{i}^{(0)},\psi_s\left(h_{i}^{(n)}\right)\right)+\lambda_p\sum_{t=0}^{n-1}\mathcal{L}_{p}\left(H^{(t)}_{Q_i}, \hat{H}^{(t)}_{Q_i}\right), \label{eq-loss}
\end{equation}
where $\hat{H}^{(t)}_{Q_i} = \psi_p^{(t+1)}\Big(h_{i}^{(t+1)}, H_{\mathcal{N}_i}^{(t+1)}\Big)$ denotes the reconstructed neighborhood set of node features based on the sampling network shown in   Fig.~\ref{fig: appd_model}\textbf{b}. Here, $Q_i\subset\mathcal{N}_{i}$ denotes the set of $q$ samples of neighborhood nodes for $v_i$; $\lambda_s$ and $\lambda_p$ are the weighting coefficients; and $\mathcal{L}_s$ stands for the reconstruction error of the node features, given by
\begin{equation}
    \mathcal{L}_{s}\left(h_{i}^{(0)},\psi_s\left(h_{i}^{(n)}\right)\right)=\left\|h_{i}^{(0)}-\psi_s\left(h_{i}^{(n)}\right)\right\|^2_2.
\end{equation}
In Eq. \eqref{eq-loss}, $\mathcal{L}_p$ is the loss function used to measure the reconstruction of the neighborhood set of node features $H^{(t)}_{Q_i}$. 
Inspired by~\cite{tang2022graph}, we evaluate this loss function by a Monte Carlo method. 
Specifically, for node $v_i$, the distribution of its neighborhood information can be empirically represented by $\mathcal{P}^{(t)}_{i}$ defined as follows:
\begin{equation}
    \mathcal{P}^{(t)}_{i} = \sum_{v_j \in \mathcal{N}_{i}}\delta_{h_{j}}^{(t)},
\end{equation}
where $\delta_{h_{j}}^{(t)}$ denotes the Dirac delta function. 
Here, we adopt the 2-Wasserstein distance, which measures the similarity between two distributions, to construct the loss \cite{tang2022graph}, expressed as
\begin{equation}
    \mathcal{L}_{p}\left(H_{\mathcal{N}_{i}}^{(t)}, \hat{H}^{(t)}_{Q_i}\right)=
    \mathcal{W}_2^2\left(\mathcal{P}^{(t)}_{i}, \hat{H}^{(t)}_{Q_i}\right).
    \label{eq:lp}
\end{equation}
In our experiments, we fix $q=10$ based on a Hungarian matching, which avoids heavy computational overhead meanwhile retaining accuracy, when evaluating Eq. \eqref{eq:lp} during training.


    \label{eq:surrogated-loss}

\paragraph{Classifier model}
The classifier model is constructed by adding a pooling layer and a softmax module after the encoder of the pre-trained model (see   Fig.~\ref{fig: appd_model}\textbf{c}).
The pooling layer is applied to the embedding of the pre-trained encoder to generate an overall feature vector $h_g$ that can be represented by a pooling function given by $h_g={\rm{Pool}}(h_0^{(0)},h_1^{(0)},...h_N^{(0)},\dots, h_N^{(n)})$, where $n$ is the number of convolution layer and $N$ is the number of nodes in graph.
The softmax module in the classifier model ensures that the output for each candidate material through the model is a probability in the range of $[0, 1]$, representing the likelihood of the candidate material being an altermagnetic material.

\subsection*{Implementation details}

\paragraph{The pre-training model}
To extract the crystal embeddings of the candidate materials, we employ a graph convolution neural network as an encoder, which consists of 3 graph convolution layers\footnote{We refer the code of CGCNN~\cite{xie2018crystal}  to construct the encoder module, whose github link is \url{https://github.com/txie-93/cgcnn}  and commit ID is ``f42ab233c4ee0c416879d6bc2d22a264418413ad''.}.
For classification, we utilize a pooling layer and a multilayer perceptron as the projection head, comprising two fully connected layers with a ReLU activation layer and a Dropout layer\footnote{We refer the code of NWR-GAE~\cite{tang2022graph} to construct the auto-decoder module and compute loss, whose github link is \url{https://github.com/mtang724/NWR-GAE} and commit ID is ``e9aee57a009f6c2150cf68fc173c2af3094a7205''.}
In terms of optimization, we use the Adam optimizer with a learning rate of 0.001. Additionally, we implement MultiStepLR with milestones set at 100. 
To ensure stability during training, we train our classifier model with a dropout rate of 0.25. 
This is done using a batch size of 512 and training over 500 epochs on 2 NVIDIA A100 GPUs.
We then label these materials whose probabilities exceed 0.9 as potential altermagnetic candidates in all experiments.
To train and evaluate our model efficiently, we leverage the distributed deep learning framework Accelerate. More details of the model training are discussed in {Supplementary Appendix Note A}. 
Given this specific network architecture, we can complete the training and evaluation process of the classifier model in less than 1.5 hours on our datasets. This significantly improves the overall efficiency of our proposed workflow.

\paragraph{The first-principles electronic calculation}
The first-principles electronic structure calculations were performed in the framework of density functional theory~(DFT) using the Vienna Abinitio Simulation Package (VASP)~\cite{PhysRevB.54.11169}. 
The generalized gradient approximation~(GGA) of the Perdew-Burke-Ernzerhof (PBE) type was adopted for the exchange-correlation functional~\cite{PhysRevLett.77.3865}. 
The projector augmented wave~(PAW) method was employed to describe the interactions between valence electrons and nuclei~\cite{PhysRevB.50.17953}. 
To account for the correlation effects of 3d and 4f orbitals, we performed $\rm GGA+U$ calculations by using the simplified rotationally invariant version introduced by Dudarev \textit{et al.}~\cite{PhysRevB.57.1505}. The detail information for DFT calculations is given in {Supplement Appendix Note D}.

\section*{SUPPLEMENTARY DATA}
Supplementary data are available at NSR online.
The crystal data are available from the Materials Project database via the web interface at \url{https://materialsproject.org} or the API at \url{https://api.materialsproject.org}. 

All the source codes to reproduce the results in this study are available on GitHub at \url{https://github.com/zfgao66/MatAltMag}.
We rely on PyTorch~(\url{https://pytorch.org}) for deep model training.
We use specialized tools for the Vienna Abinitio Simulation Package~(\url{https://www.vasp.at/}).
The code of the pre-training model for crystal materials and the pre-trained neural-network weights are available on GitHub at \url{https://github.com/zfgao66/MatAltMag}.

\section*{ACKNOWLEDGEMENTS}
The authors would like to thank Prof. Hongteng Xu for the discussion of the graph neural network architecture. Computational resources have been provided by the Physical Laboratory of High Performance Computing at Renmin University of China.

\section*{FUNDING}
The work is supported by the National Natural Science Foundation of China (No. 62276269, No. 92270118, No.62476278, No.12204533, and No.11934020) and the Beijing Natural Science Foundation (No. 1232009). 

\section*{AUTHOR CONTRIBUTIONS}  
Z.F.G., H.S., P.J.G., and Z.Y.L. contributed to the ideation and design of the research; Z.F.G., S.Q., B.Z., and P.J.G. performed the research; Z.F.G., H.S., P.J.G., and Z.Y.L. wrote and edited the paper; all authors contributed to the research discussions.

\noindent\emph{\textbf{Conflict of interest statements.}} None declared.

\bibliographystyle{nsr}
\bibliography{references}

\end{document}


\title{\textbf{Supplementary Information} for: \\ AI-accelerated Discovery of Altermagnetic Materials}

\author[1,2,$\dag$]{Ze-Feng Gao}
\author[2,$\dag$]{Shuai Qu}
\author[1,$\dag$]{Bocheng Zeng}
\author[3]{Yang Liu} 
\author[1]{Ji-Rong Wen} 
\author[1,*]{\\Hao Sun}
\author[2,*]{Peng-Jie Guo}
\author[2,*]{Zhong-Yi Lu}

\affil[1]{\small Gaoling School of Artificial Intelligence, Renmin University of China, Beijing, China}
\affil[2]{\small Department of Physics, Renmin University of China, Beijing, China}
\affil[3]{\small School of Engineering Science, University of Chinese Academy of Sciences, Beijing, China\vspace{18pt}} 
\affil[$\dag$]{Equally contributed\vspace{6pt}}
\affil[*]{Corresponding authors\vspace{12pt}}

\date{}

\maketitle

\vspace{-36pt}

{\footnotesize
\tableofcontents
}

\vspace{24pt}

\noindent This supplementary document provides a detailed description of the proposed pre-trained model, dataset statistics, hyperparameter value, and details of altermagnetic materials confirmed by electronic structure calculations.


\section{Addition information for pre-trained model}
\paragraph{Model architecture.}
We have two models in total, which are the auto-encoder model and the classifier model. Fig. 2 in the Main text shows the overall architecture and details of models. We pre-process the crystals and construct a graph for each crystal according to its structure and atoms. In Fig. 2 in the Main text, $\bm{\mu}_v$ stands for features from the crystal graph.


\paragraph{Workflow for symmetry analysis.}
Fig.~\ref{fig: symmetry restriction} has shown the workflow for symmetry analysis, through this screening process, we can build the pre-training dataset, fine-tuning dataset, and candidate materials dataset.
\begin{figure}[t]
    \centering
    \includegraphics[width=\textwidth]{section/image/SI-fig/Figure-S.1-sysmetry-procedure.pdf}
    \caption{
    \textbf{The workflow for screening altermagnetic materials based on symmetry analysis.} Through this screening process, we can build datasets for pre-training and fine-tuning, as well as a candidate set of possible altermagnetic materials.
    }
    \label{fig: symmetry restriction}
\end{figure}

\paragraph{Visulization of the learned crystal embeddings.}
We perform visualization for the learned crystal embeddings of the candidate materials dataset, which we projected into 20 principal components with the PCA technique. We present the pair-wise visualization of three or two selected principal components of the learned crystal embeddings. Moreover, we also perform the T-SNE procedure to plot the 20 principal components to reveal the obvious clustering phenomena.

\begin{figure}[ht]
    \centering
    \includegraphics[width=0.85\textwidth]{section/image/SI-fig/Figure-S.2-pca.pdf}
    \caption{\textbf{Visualization of the learned crystal embeddings of the candidate materials, projected into 20 principal components by PCA.} 
    {\bf{a}}--{\bf{c}}, Pair-wise visualization of three or two selected principal components of the learned crystal embeddings. 
    {\bf{d}}, The t-distributed stochastic neighbor embedding (t-SNE) plot of the 20 principal components, reveals obvious clustering phenomena.
    }
    \label{fig:visual_encoder}
\end{figure}

\paragraph{Learning curve of auto-encoder.}
We pre-train our auto-encoder model for only 10 epochs. To show the changing trend of pre-train loss, we choose the number of batches as the $x$-coordinate instead of epochs in Fig.~\ref{fig:learning_curve}.

\begin{figure}[ht]
    \centering
    \includegraphics[width=0.6\textwidth]{section/image/SI-fig/Figure-S.3-pretrain_loss.pdf}
    \caption{
    \textbf{Learning curve of the auto-encoder model with varying learning rate.}
    The $x$-coordinate is the number of batches while the $y$-coordinate is the MSE loss.}
    \label{fig:learning_curve}
\end{figure}

\paragraph{Details for pre-processing.}
\label{appdix-crystal-graph}
The 3-dim crystal structure is straightforward and intuitive, but it is awkward to deal with in code. Thus, the crystal graph arises from the 3-dim crystal structure.
The part of negligible information is not significant, such as the weak bonds. The nodes and edges represent the atoms and bonds, respectively. The nodes are connected with edges. We still retain the important information of atoms and bonds.
The magnetic atoms are extracted with the corresponding weight function.
We are concerned with the material information of altermagnetic, which is critical and indispensable.

The feature vector $h$ is used throughout each step. In order to make it easier to understand, we are going to expand on it. The feature vector $h$ is analogous to the occupation number representation in quantum mechanics with the range from unoccupied 0 to occupied 1. The occupancy probability given by the subsequent iterations is still between 0 and 1 and it is the superposition of the wannier wave functions of the unoccupied state and occupied state. The element order information in the feature vector $h$ is read from the structure file, and the order is fixed.
There is also a significant eigenvector $z$, which is effective in complementing the information about the interaction strength of the bonds. The initial eigenvector $u(i,j)_k$ has only two values of 0 and 1, without bond 0 and with bond 1. However, the strong and weak information will be manifested by the relative magnitude with the range from 0 and 1 in the subsequent iterations. The stronger bonds will be close to 1, and the weaker bonds will be close to 0. 
An example of pre-processing for FeB to graph representation is shown in Fig.~\ref{fig: app-crystal-graph}. 

\begin{figure}[h]
    \centering
    \includegraphics[width=\textwidth]{section/image/SI-fig/Figure-S.4-FeB.pdf}
   \caption{
    \textbf{An example of pre-processing for crystal structure to a crystal graph representation.}
    The green nodes denote the B atoms, and the brown nodes denote the Fe atoms.
    The Fe1 node connects B1, B2, B3, and B4 nodes by 4 edges. It is pronounced that the Fe2 node connects the B1 and B4 nodes.
    The Fe nodes are consistent with each other in the 4 corners with periodic boundary conditions. Thus, the Fe2 node connects B1, B2, B3, and B4 nodes and the Fe1 node. Furthermore, the top B2 node connects to the bottom B1 node, and the left B3 node connects to the right B4 node.
    }
    \label{fig: app-crystal-graph}
\end{figure}

\paragraph{Hyperparameter value.}
\begin{table}[h]
\centering
\caption{\bf{Hyperparameter value obtained by grid search.}}
\begin{tabular}{llll}
\toprule
Hyperparameter   & Auto-encoder & Classifier & Description                                                                     \\ \hline
Epochs           & 10           & 500        & Training epoch                                                                  \\
Learning rate    & 1.0e-3       & 1.0e-3     & Learning rate for optimizing neural network                                     \\
Batch size       & 64           & 512         & Number of input token per batch                                                 \\
Hidden dimension & 512          & 512        & Size of dimensions in the graph convolution layer                                                       \\
Sample size      & 10           & -          & Number of samples during the reconstruction \\
Radius           & 20           & 20         & Neighbor distance of crystal atoms                                              \\
Drop rate        & -            & 0.25       & Dropout rate for Classifier                                                     \\ \bottomrule
\end{tabular}
\end{table}

In the pre-processing procedure, the crystal structure files (.cif) are transformed into the crystal graph data which can be fed into the model directly. Nodes of the graph are represented by atoms naturally and the atoms will be regarded as the neighbors of another atom if they are in the unit cell, out to a distance of radius. In the pre-training procedure, using the pre-training dataset consisting of 68,116 materials, we pre-train our auto-encoder model for 10 epochs with a batch size of 64 on 2 NVIDIA A100 GPUs, which takes about 2 days. Once pre-training is over, we load the weights of the encoder module into the classifier model and begin training for 500 epochs with a batch size of 512 and a drop rate of 0.25. The learning rate of both models is 0.001 and the hidden dimension of the encoder module is 512.

\section{Addition altermagnetic materials confirmed by electronic structure calculations}

For the 19 materials (NdRuO$_3$, NaFeO$_2$(31), MnO$_2$(62), MnO$_2$(87), CaLaCr$_2$O$_6$, Ca$_3Cr_2$O$_7$, CaLaFeAgO$_6$, ZrCrO$_3$, ZrMnO$_3$, VF$_3$, CrF$_3$, NiF$_3$, CaMnN$_2$, Ir$_5$B$_2$Mg$_2$Fe, Ir$_5$B$_2$Mg$_2$Mn, Ir$_5$B$_2$Mg$_2$Ni, Ir$_5$B$_2$Sc$_2$V, Ir$_5$B$_2$Sc$_2$Mn), we determine their magnetic ground states by calculating the energies of their different magnetic structures under different correlated interactions U. Then, symmetry analysis is used to determine that these magnetic ground states are altermagnetic states. Furthermore, symmetry analysis is also used to determine whether these altermagnetic materials are $d$-wave, $g$-wave, or $i$-wave. Finally, the electronic band structures are used to demonstrate our symmetry analysis. All calculated results are shown in Figs.~\ref{fig: NdRuO3}--\ref{fig: VF3}.

\begin{figure}[h]
    \centering
    \includegraphics[width=0.99\textwidth]{section/image/SI-fig/Figure-S.5-NdRuO3.pdf}
    \caption{{
    \textbf{The crystal and electronic structure of the altermagnet NdRuO$_3$.}
    {\bf{a}}, The side view of the magnetic primitive cell of the altermagnetic NdRuO$_3$.
    {\bf{b}}, The top view of the magnetic primitive cell of the altermagnetic NdRuO$_3$.
    The arrows represent the magnetic moments of Nd and Ru. 
    {\bf{c}}, The Brillouin zone (BZ) with high-symmetry points and lines of altermagnetic NdRuO$_3$.
    {\bf{d}}, The anisotropic spin-charge density deriving from an anisotropic crystal field. 
    {\bf{e}} and {\bf{f}} are the electronic band structure of NdRuO$_3$ without SOC along different high-symmetry directions. 
    The red and blue lines represent the spin-up and spin-down energy bands, respectively.
    The electronic structure is calculated under correlation interaction $\rm{U= 7~eV}$ and $\rm{U= 2~eV}$ for 4f orbits of Nd and 4d obits of Ru.}}
    \label{fig: NdRuO3}
\end{figure}


\begin{figure}[h]
    \centering
    \includegraphics[width=\textwidth]{section/image/SI-fig/Figure-S.6-NaFeO2(33).pdf}
    \caption{
    \textbf{Altermagnetic NaFeO$_2$(33) is confirmed by electronic structure calculations.}
    {\bf{a}}, The crystal structures of the NaFeO$_2$(33) with FM.
    {\bf{b}}, The crystal structures of the NaFeO$_2$(33) with AFM1.
    {\bf{c}}, The crystal structures of the NaFeO$_2$(33) with AFM2.
    {\bf{d}}, The crystal structures of the NaFeO$_2$(33) with AFM3.
    The arrows represent the magnetic moments of Fe. 
    {\bf{e}}, The relative energy of four magnetic states with the variation of correlation interaction U. 
    {\bf{f}}, The electronic band structure of NaFeO$_2$(33) along non-high-symmetry directions without SOC. 
    The red and blue lines represent the spin-up and spin-down energy bands, respectively.
    The electronic structure is calculated under correlation interaction $\rm{U= 4~eV}$.
    }
    \label{fig: NaFeO2}
\end{figure}


\begin{figure}[h]
    \centering
    \includegraphics[width=\textwidth]{section/image/SI-fig/Figure-S.7-MnO2-62.pdf}
    \caption{
    \textbf{Altermagnetic MnO$_2$(62) is confirmed by electronic structure calculations.}
    {\bf{a}}, The crystal structures of the MnO$_2$(62) with FM.
    {\bf{b}}, The crystal structures of the MnO$_2$(62) with AFM1.
    {\bf{c}}, The crystal structures of the MnO$_2$(62) with AFM2.
    {\bf{d}}, The crystal structures of the MnO$_2$(62) with AFM3.
    The arrows represent magnetic moments of Mn. 
    {\bf{e}}, The relative energy of four magnetic states with the variation of correlation interaction U. 
    {\bf{f}}, The electronic band structure along non-high-symmetry directions without SOC. 
    The red and blue lines represent the spin-up and spin-down energy bands, respectively.
    The electronic structure is calculated under correlation interaction $\rm{U= 4~eV}$.
    }
    \label{fig: MnO2-62}
\end{figure}


\begin{figure}[h]
    \centering
    \includegraphics[width=\textwidth]{section/image/SI-fig/Figure-S.8-MnO2-87.pdf}
    \caption{
    \textbf{Altermagnetic MnO$_2$(87) is confirmed by electronic structure calculations.}
    {\bf{a}}, The crystal structures of the MnO$_2$(87) with FM.
    {\bf{b}}, The crystal structures of the MnO$_2$(87) with AFM1.
    {\bf{c}}, The crystal structures of the MnO$_2$(87) with AFM2.
    {\bf{d}}, The crystal structures of the MnO$_2$(87) with AFM3.
    The arrows represent magnetic moments of Mn. 
    {\bf{e}}, The relative energy of four magnetic states with the variation of correlation interaction U. 
    {\bf{f}}, The electronic band structure along non-high-symmetry directions without SOC. 
    The red and blue lines represent the spin-up and spin-down energy bands, respectively.
    The electronic structure is calculated under correlation interaction $\rm{U= 4~eV}$.
    }
    \label{fig: MnO2-87}
\end{figure}

\begin{figure}[h]
    \centering
    \includegraphics[width=\textwidth]{section/image/SI-fig/Figure-S.9-CaLaCr2O6.pdf}
    \caption{
    \textbf{Altermagnetic CaLaCr$_2$O$_6$ is confirmed by electronic structure calculations.}
    {\bf{a}}, The crystal structures of the CaLaCr$_2$O$_6$ with FM.
    {\bf{b}}, The crystal structures of the CaLaCr$_2$O$_6$ with AFM1.
    {\bf{c}}, The crystal structures of the CaLaCr$_2$O$_6$ with AFM2.
    {\bf{d}}, The crystal structures of the CaLaCr$_2$O$_6$ with AFM3.
    The arrows represent magnetic moments of Cr. 
    {\bf{e}}, The relative energy of four magnetic states with the variation of correlation interaction U. 
    {\bf{f}}, The electronic band structure along non-high-symmetry directions without SOC. 
    The red and blue lines represent the spin-up and spin-down energy bands, respectively.
    The electronic structure is calculated under correlation interaction $\rm{U= 4~eV}$.
    }
    \label{fig: CaLaCr2O6}
\end{figure}


\begin{figure}[h]
    \centering
    \includegraphics[width=\textwidth]{section/image/SI-fig/Figure-S.10-Ca3Cr2O7.pdf}
    \caption{
    \textbf{Altermagnetic Ca$_3$Cr$_2$O$_7$ is confirmed by electronic structure calculations.}
    {\bf{a}}, The crystal structures of the Ca$_3$Cr$_2$O$_7$ with FM.
    {\bf{b}}, The crystal structures of the Ca$_3$Cr$_2$O$_7$ with AFM1.
    {\bf{c}}, The crystal structures of the Ca$_3$Cr$_2$O$_7$ with AFM2.
    {\bf{d}}, The crystal structures of the Ca$_3$Cr$_2$O$_7$ with AFM3.
    The arrows represent magnetic moments of Cr. 
    {\bf{e}}, The relative energy of four magnetic states with the variation of correlation interaction U. 
    {\bf{f}}, The electronic band structure along non-high-symmetry directions without SOC. 
    The red and blue lines represent the spin-up and spin-down energy bands, respectively.
    The electronic structure is calculated under correlation interaction $\rm{U= 4~eV}$.
    }
    \label{fig: Ca3Cr2O7}
\end{figure}


\begin{figure}[h]
    \centering
    \includegraphics[width=\textwidth]{section/image/SI-fig/Figure-S.11-CaLaFeAgO6.pdf}
    \caption{
    \textbf{The crystal and electronic structure of the altermagnetic CaLaFeAgO$_6$.}
    {\bf{a}}, The crystal primitive cell of altermagnetic CaLaFeAgO$_6$.
    {\bf{b}}, The crystal structures of the CaLaFeAgO$_6$ with FM.
    {\bf{c}}, The crystal structures of the CaLaFeAgO$_6$ with AFM1.
    {\bf{d}}, The crystal structures of the CaLaFeAgO$_6$ with AFM2.
    {\bf{e}}, The crystal structures of the CaLaFeAgO$_6$ with AFM3.
    The arrows represent magnetic moments of Fe. 
    {\bf{f}}, The Brillouin zone (BZ) with high-symmetry points and lines of altermagnetic CaLaFeAgO$_6$.
    {\bf{g}}, The relative energy of four magnetic states with the variation of correlation interaction U. 
    {\bf{h}}, The electronic band structure along high-symmetry directions without SOC. 
    {\bf{i}}, The electronic band structure along non-high-symmetry directions without SOC.
    The red and blue lines represent the spin-up and spin-down energy bands, respectively.
    The electronic structure is calculated under correlation interaction $\rm{U= 4~eV}$.
    }
    \label{fig: CaLaFeAgO6}
\end{figure}


\begin{figure}[h]
    \centering
    \includegraphics[width=\textwidth]{section/image/SI-fig/Figure-S.12-ZrCrO3.pdf}
    \caption{
    \textbf{Altermagnetic ZrCrO$_3$ is confirmed by electronic structure calculations.}
    {\bf{a}}, The crystal structures of the ZrCrO$_3$ with FM.
    {\bf{b}}, The crystal structures of the ZrCrO$_3$ with AFM1.
    {\bf{c}}, The crystal structures of the ZrCrO$_3$ with AFM2.
    {\bf{d}}, The crystal structures of the ZrCrO$_3$ with AFM3.
    The arrows represent magnetic moments of Cr. 
    {\bf{e}}, The relative energy of four magnetic states with the variation of correlation interaction U. 
    {\bf{f}}, The electronic band structure along non-high-symmetry directions without SOC. 
    The red and blue lines represent the spin-up and spin-down energy bands, respectively.
    The electronic structure is calculated under correlation interaction $\rm{U= 4~eV}$.
    }
    \label{fig: ZrCrO3}
\end{figure}


\begin{figure}[h]
    \centering
    \includegraphics[width=\textwidth]{section/image/SI-fig/Figure-S.13-ZrMnO3.pdf}
    \caption{
    \textbf{Altermagnetic ZrMnO$_3$ is confirmed by electronic structure calculations.}
    {\bf{a}}, The crystal structures of the ZrMnO$_3$ with FM.
    {\bf{b}}, The crystal structures of the ZrMnO$_3$ with AFM1.
    {\bf{c}}, The crystal structures of the ZrMnO$_3$ with AFM2.
    {\bf{d}}, The crystal structures of the ZrMnO$_3$ with AFM3.
    The arrows represent magnetic moments of Mn. 
    {\bf{e}}, The relative energy of four magnetic states with the variation of correlation interaction U. 
    {\bf{f}}, The electronic band structure along non-high-symmetry directions without SOC. 
    The red and blue lines represent the spin-up and spin-down energy bands, respectively.
    The electronic structure is calculated under correlation interaction $\rm{U= 4~eV}$.
    }
    \label{fig: ZrMnO3}
\end{figure}


\begin{figure}[h]
    \centering
    \includegraphics[width=\textwidth]{section/image/SI-fig/Figure-S.14-CaMnN2.pdf}
    \caption{
    \textbf{The crystal and electronic structure of the altermagnetic CaMnN$_2$.}
    {\bf{a}}, The crystal primitive cell of altermagnetic CaMnN$_2$.
    {\bf{b}}, The crystal structures of the CaMnN$_2$ with FM.
    {\bf{c}}, The crystal structures of the CaMnN$_2$ with AFM1.
    {\bf{d}}, The crystal structures of the CaMnN$_2$ with AFM2.
    {\bf{e}}, The crystal structures of the CaMnN$_2$ with AFM3.
    The arrows represent magnetic moments of Mn. 
    {\bf{f}}, The Brillouin zone (BZ) with high-symmetry points and lines of altermagnetic CaMnN$_2$.
    {\bf{g}}, The relative energy of four magnetic states with the variation of correlation interaction U. 
    {\bf{h}}, The electronic band structure along high-symmetry directions without SOC. 
    {\bf{i}}, The electronic band structure along non-high-symmetry directions without SOC.
    The red and blue lines represent the spin-up and spin-down energy bands, respectively.
    The electronic structure is calculated under correlation interaction $\rm{U= 4~eV}$.
    }
    \label{fig: CaMnN2}
\end{figure}

\begin{figure}[h!]
    \centering
    \includegraphics[width=0.99\textwidth]{section/image/SI-fig/Figure-S.15-MnO.pdf}
    \caption{
    \textbf{The crystal and electronic structure of the altermagnet MnO.}
    {\bf{a}}, The crystal primitive cell of altermagnetic MnO.
    {\bf{b}}--{\bf{e}}, The four significant magnetic structures MnO. The arrows represent magnetic moments of $Mn$. 
    {\bf{f}}, The Brillouin zone (BZ) with high-symmetry points and lines of altermagnetic MnO.
    {\bf{g}}, The relative energy of four significant magnetic states with the variation of correlation interaction U. 
    {\bf{h}} and {\bf{i}}, The electronic band structure of MnO along high-symmetry directions and non-high-symmetry directions without spin-orbit coupling~(SOC), respectively. 
    The red and blue lines represent the spin-up and spin-down energy bands, respectively.
    The electronic structure is calculated under correlation interaction $\rm{U= 4~eV}$.
    }
    \label{fig: MnO}
\end{figure}

\begin{figure}[h]
    \centering
    \includegraphics[width=0.95\textwidth]{section/image/SI-fig/Figure-S.16-IrB-set.pdf}
    \caption{ 
    \textbf{Altermagnetic materials are confirmed by electronic structure calculations.}
    {\bf{a}}, The crystal primitive cell. {\bf{b}}--{\bf{e}}, The crystal structures of the Ir$_5$B$_2$Y$_2$X~(Y=Mg,Sc,X=Fe,Mn,Ni,V) with FM, AFM1, AFM2 and AFM3 magnetic structures, respectively. {\bf{f}}--{\bf{j}}, Relative energy of four significant magnetic states with the variation of correlation interaction U of Ir$_5$B$_2$Mg$_2$Fe, Ir$_5$B$_2$Mg$_2$Mn, Ir$_5$B$_2$Mg$_2$Ni, Ir$_5$B$_2$Sc$_2$V and Ir$_5$B$_2$Sc$_2$Mn. {\bf{k}}--{\bf{o}}, The electronic band structure of altermagnetic Ir$_5$B$_2$Y$_2$X along non-high-symmetry directions without SOC. The red and blue lines represent spin-up and spin-down energy bands, respectively.
    }
    \label{fig: IrB}
\end{figure}

\begin{figure}[h]
    \centering
    \includegraphics[width=0.95\textwidth]{section/image/SI-fig/Figure-S.17-VF-set.pdf}
    \caption{  
    \textbf{Altermagnetic materials are confirmed by electronic structure calculations.}
    {\bf{a}}, The crystal primitive cell and Brillouin Zone~(BZ).
    {\bf{b}}--{\bf{e}}, The crystal structures of XF$_3$~(X=V,Cr,Ni) with FM, AFM1, AFM2 and AFM3 magnetic structures, respectively. 
    {\bf{f}}--{\bf{j}}, Relative energy of four significant magnetic states with the variation of correlation interaction U of VF$_3$, CrF$_3$, and NiF$_3$. 
    {\bf{k}}--{\bf{o}}, The electronic band structure of altermagnetic XF$_3$ along non-high-symmetry directions without SOC.
    The red and blue lines represent spin-up and spin-down energy bands, respectively.
    The electronic structure is calculated under correlation interaction $\rm{U= 4~eV}$.
    }
    \label{fig: VF3}
\end{figure}

\begin{table}[t!]
\centering
\caption{
{\bf{The band gap of the 34 altermagentic insulator materials under the LDA+U framework.}}
}
\footnotesize
\begin{tabular}{ccc}
\toprule
{Number} & {Materials}  & {Gap/eV}  \\ \hline
{1} &{$\text{Fe}\text{H}\text{O}_2(31)$}           & {1.94}          \\
{2} &{$\text{Na}\text{Fe}\text{O}_2~(33)$}        & {3.18}        \\
{3} &{$\text{Na}\text{Fe}\text{O}_2~(92)$}        & {3.11}       \\
{4} &{$\text{Mn}\text{O}_2~(62)$}      & {1.60}        \\
{5} &{$\text{Mn}\text{O}_2~(87)$}      & {1.73}       \\
{6} &{$\text{Ca}_3\text{Cr}_2\text{O}_7$}   & {1.34}          \\
{7} &{$\text{Zr}\text{Cr}\text{O}_3$}          & {1.86}            \\
{8} &{$\text{Zr}\text{Mn}\text{O}_3$}        & {3.47}              \\
{9} &{$\text{V}\text{F}_3$}       & {2.87}            \\ 
{10} &{$\text{Cr}\text{F}_3$}       & {4.13}            \\ 
{11} &{$\text{Mn}\text{O}$}      & {0.65}       \\
{12} &{$\text{Ca}\text{Mn}\text{N}_2$}      & {0.95} \\
{13} &{$\text{V}\text{F}_4$} &{3.69}\\
{14} &{$\text{Ca}_2\text{Co}\text{Te}\text{O}_6$} &{2.04}\\
{15} &{$\text{Ni}\text{F}_2$} &{4.70}\\
{16} &{$\text{Ba}_2\text{Fe}\text{Ge}_2\text{O}_7$} &{1.85}\\
{17} &{$\text{Ba}_2\text{Co}\text{Si}_2\text{O}_7$} &{3.15}\\
{18} &{$\text{Sr}_2\text{Co}\text{Ge}_2\text{O}_7$} &{1.93}\\
{19} &{$\text{Fe}\text{H}\text{O}_2(19)$} &{2.26}\\
{20} &{$\text{Ca}\text{Mn}\text{O}_3$} &{1.21}\\
{21} &{$\text{Ca}\text{V}\text{O}_3$} &{1.11}\\
{22} &{$\text{La}\text{Fe}\text{O}_3$} &{2.45}\\
{23} &{$\text{La}\text{V}\text{O}_3$} &{2.22}\\
{24} &{$\text{Mn}\text{Se}\text{O}_4$} &{2.74}\\
{25} &{$\text{Na}\text{Pr}_2\text{Os}\text{O}_6$} &{0.45}\\
{26} &{$\text{Na}\text{Pr}_2\text{Ru}\text{O}_6$} &{0.76}\\
{27} &{$\text{Nd}\text{Rh}\text{O}_3$} &{1.47}\\
{28} &{$\text{Pr}\text{Ru}\text{O}_3$} &{0.18}\\
{29} &{$\text{Sc}\text{V}\text{O}_3$} &{2.30}\\
{30} &{$\text{Sm}\text{Rh}\text{O}_3$} &{1.18}\\
{31} &{$\text{Ca}\text{La}\text{Cr}\text{Mo}\text{O}_6$} &{0.83}\\
{32} &{$\text{La}2\text{Mn}\text{Rh}\text{O}_6$} &{0.44}\\
{33} &{$\text{Li}\text{Fe}\text{F}_4$} &{3.60}\\
{34} &{$\text{Li}\text{Fe}_2\text{F}_6$} &{1.55}\\
\bottomrule
\end{tabular}
\end{table}
\clearpage
\newpage

\section{Addition discussion for proposed AI search engine}
{\paragraph{Intrinsic errors and computational costs associated with the proposed algorithm.}
About the intrinsic errors and computational costs associated with the algorithm, we provide a detailed breakdown of each computational step:

\vspace{-3pt}
\begin{itemize}

\item[(1)] \textbf{Dataset Collection} -- \textit{collecting the pre-training dataset, fine-tuning dataset, and candidate materials through high-throughput screening}: This step requires relatively less efforts since data can be fetched using the API provided by the Material Project. 

\vspace{-3pt}

\item[(2)] \textbf{Model Pre-training} -- \textit{establishing and pre-training the GNN model using the pre-training dataset}: This step allows to learn the intrinsic features of the crystal structure of magnetic materials. The computational cost of this step is mainly related to the volume of pre-training data, the size of the GNN model, and the number of training iterations. In our experiments, the training was conducted using four Nvidia A100 GPUs, taking approximately 35 hours to converge.

\vspace{-3pt}

\item[(3)] \textbf{Classifier Fine-tuning} -- \textit{fine-tuning the classifier model using the fine-tuning dataset}: This step allows the model to predict altermagnetism probability of a given material candidate. Possible intrinsic errors here may arise from differences between the estimated distribution of the positive samples and the true distribution, which can lead to biased prediction by the model. This can be mitigated by iteratively adding more positive samples (e.g., once a new altermagnetic material is discovered, it is added to the list of positive samples to refine the classifier model). Since there are quite a limited number of positive samples, the training effort in this step typically takes about 2--3 hours on one Nvidia A100 GPU.

\vspace{-3pt}

\item[(4)] \textbf{Material Verification} -- \textit{conducting DFT electronic structure calculations on the predicted candidate materials to verify their altermagnetic properties}: This step involves determining the magnetic ground state and band structure, whose computational cost depends on factors such as the number of atoms in the primitive cell. Typically, the calculation of each material for verification takes about 12--48 hours. Note that the calculations were conducted on a server equipped with 64-core Intel Xeon Gold 6438 processors with parallel computing.

\end{itemize}

\vspace{-3pt}

\paragraph{AI component \emph{v.s.} simply searching based on physics.}
The “brute-force” search based on physics involves guessing potential crystal materials by experts followed by DFT calculations for verification, which are not very quantifiable in terms of accuracy and efficiency. This conventional strategy is undesirable given tens of thousands of candidate materials. In the contrary, AI models have two advantages: (1) they rely less on experts’ experience, which leverage known positive samples for training and prediction; (2) through 3 active learning, the predicted positive samples can be further added to the training data to iteratively improve the accuracy of the model’s prediction.

\paragraph{The convergence of the proposed AI search engine.}
In our experiments, after the four rounds of iteration, the candidate materials predicted by the model predominantly consist of materials with a huge number of atoms in the unit cell\footnote{We consider crystalline materials with unit cells containing 40 or more atoms as materials with a huge number of atoms within the unit cell.}). In other words, we can no longer determine whether these materials exhibit altermagnetic properties (although they may possess them) through finite DFT calculations. Therefore, after four iterations, we concluded that the model had converged and could not predict additional altermagnetic materials.

\paragraph{The few-shot learning ability of proposed AI search engine.}
The AI search engine can obtain predictability by fine-tuning procedures. 
The model performance was positive relative to the number of positive samples.
In our experiment, we predicted 25 altermagnetic materials with only 14 positive samples. This result shows the few-shot learning ability of the AI search engine.
Moreover, we added the rest 134 positive samples into the fine-tuned dataset, and then the model performance was significantly improved.
Note that the model’s performance is not directly related to spin patterns, and the model cannot directly distinguish between AFM1, AFM2, and AFM3 spin patterns. 

\paragraph{Accuracy of the classifier model.}
Based on our current results, the accuracy of finding altermagnetic materials using the AI search engine is around 31\% (e.g., 50 new altermagnetic materials were found, out of 161 candidate materials predicted by AI which were calculated by DFT). The accuracy of the classifier model might be improved by adding more positive samples. In contrast, when we evaluated 10 materials recommended by human experts, we could not discover any new altermagnetic properties. Such a process depends on the expertise of experts and may yield easily a biased estimation.

\section{Addition information for DFT calculations}
{We use the first principle of electronic calculations to verify the predicted materials by the AI search engine whether it is altermagnetic. Table~\ref{dft-details} lists the hyper-parameters for the density wave functions (DFT) in our main experiments.}

\begin{table}[htbp]
\centering
\caption{{The hyper-parameter for DFT calculations.}}
\footnotesize
\begin{tabular}{lllccll}
\toprule
{{Number}} & {{Materials}}  & {{Space group}} & {{ENCUT/eV}} & {{$\bm{k}$-grid}} & {U-value/eV} & {Calculation $\mu_B$} \\ \hline
{{1}} & {{$\text{Nb}_2\text{Fe}\text{B}_2$}}    & {{$P4-mbm~(127)$}} & {{600}}     & {{$8\times8\times15$}} & {Fe:5} &{Fe:2.93}         \\
{{2}} & {{$\text{Ta}_2\text{Fe}\text{B}_2$}}    & {{$P4-mbm~(127)$}} & {{600}}     & {{$8\times8\times15$}} & {Fe:5} &{Fe:2.80}         \\
{{3}} &{{$\text{Nd}\text{B}_2\text{C}_2$}}     & {{$P4-mbm~(127)$}} & {{600}}     & {{$10\times10\times16$}} & {Nd:7} &{Nd:3.06}          \\
{{4}} &{{$\text{Mg}_2\text{Fe}\text{Ir}_5\text{B}_2$}} & {{$P4-mbm~(127)$}} & {{600}}     & {{$6\times6\times18$}} & {Fe:4}
 &{Fe:3.52}      \\
{{5}} &{{$\text{Mg}_2\text{Mn}\text{Ir}_5\text{B}_2$}} & {{$P4-mbm~(127)$}} & {{600}}     & {{$6\times6\times18$}} & {Mn:4}
 &{Mn:3.98}     \\
{{6}} &{{$\text{Mg}_2\text{Ni}\text{Ir}_5\text{B}_2$}} & {{$P4-mbm~(127)$}} & {{600}}     & {{$6\times6\times18$}} & {Ni:6.6}
 &{Ni:1.33}      \\
{{7}} &{{$\text{Sc}_2\text{V}\text{Ir}_5\text{B}_2$}}  & {{$P4-mbm~(127)$}} & {{600}}     & {{$6\times6\times18$}} & {V:3}
 &{V:1.21}       \\
{{8}} &{{$\text{Sc}_2\text{Mn}\text{Ir}_5\text{B}_2$}} & {{$P4-mbm~(127)$}} & {{600}}     & {{$6\times6\times18$}} & {Mn:4}
&{Mn:4.14}       \\
{{9}} &{{$\text{Ca}\text{La}\text{Fe}\text{Ag}\text{O}_6$}} & {{$Pc~(7)$}} & {{600}}     & {{$12\times8\times6$}}  & {Fe:4}  &{Fe:3.85}      \\
{{10}} &{{$\text{Ca}\text{La}\text{Cr}_2\text{O}_6$}} & {{$Pmn2_1~(31)$}}   & {{600}}     & {{$10\times10\times8$}}   & {Cr:4}    &{Cr:2.58}       \\
{11} &{{$\text{Ni}\text{F}_3$}}     & {{$R\bar{3}C~(167)$}}     & {{600}}     & {{$16\times16\times16$}} & {Ni:6.7} &{Ni:1.53}             \\
{12} &{{$\text{Gd}\text{B}_2\text{C}_2$}}                 & {{$P4/mbm~(127)$}}     & {{600}}     & {{$8\times8\times14$}} & {Gd:7} &{Gd:7.15} \\
{13} &{{$\text{Ho}\text{B}_2\text{C}_2$}}                 & {{$P4/mbm~(127)$}}     & {{600}}     & {{$8\times8\times14$}} & {Ho:7} &{Ho:4.09} \\
{14} &{{$\text{Lu}\text{Cr}\text{O}_3$}}     & {{$Pnma~(62)$}}     & {{600}}     & {{$10\times10\times8$}} & {Lu:7;Cr:4} &{Lu:0.00;Cr:0.97}             \\
{15} &{{$\text{Ta}\text{Co}\text{B}_2$}}                  & {{$Pnma~(62)$}}     & {{600}}    & {{$8\times16\times6$}} & {Co:3} &{Co:1.13} \\
{16} &{{$\text{Nd}\text{Ru}\text{O}_3$}}     & {{$Pnma~(62)$}}     & {{600}}     & {{$8\times8\times10$}} & {Nd:7;Ru:2} &{Nd:3.00;Ru:0.73}             \\
{17} &{{$\text{Fe}\text{H}\text{O}_2$}}      & {{$Pmn2_1~(31)$}}   & {{600}}     & {{$18\times12\times10$}} & {Fe:4} &{Fe:4.16}           \\
{18} &{{$\text{Na}\text{Fe}\text{O}_2$}}     & {{$Pna2_1~(33)$}} & {{600}}     & {{$10\times8\times10$}} & {Fe:4}  &{Fe:4.01}       \\
{19} &{{$\text{Na}\text{Fe}\text{O}_2$}}     & {{$P4_12_12~(92)$}} & {{600}}     & {{$10\times10\times8$}} & {Fe:4} &{Fe:3.98}       \\
{20} &{{$\text{Mn}\text{O}_2$}}     & {{$ Pnma~(62)$}} & {{600}}     & {{$6\times18\times12$}} & {Mn:4} &{Mn:3.04}        \\
{21} &{{$\text{Mn}\text{O}_2$}}     & {{$ I4/m~(87)$}} & {{600}}     & {{$8\times8\times4$}} & {Mn:4} &{Mn:3.03}       \\
{22} &{{$\text{Ca}_3\text{Cr}_2\text{O}_7$}} & {{$Cmc2_1~(36)$}}   & {{600}}     & {{$10\times10\times10$}}      & {Cr:4} &{Cr:2.06}         \\
{23} &{{$\text{Zr}\text{Cr}\text{O}_3$}}      & {{$Pnma~(62)$}}   & {{600}}     & {{$10\times10\times6$}}      & {Cr:4} &{Cr:3.73}         \\
{24} &{{$\text{Zr}\text{Mn}\text{O}_3$}}      & {{$R3c~(161)$}}   & {{600}}     & {{$12\times12\times12$}}       & {Mn:4} & {Mn:4.59}      \\
{25} &{{$\text{V}\text{F}_3$}}     & {{$R\bar{3}C~(167)$}}     & {{600}}     & {{$16\times16\times16$}} & {V:3} &{V:1.87}            \\ 
{26} &{{$\text{Cr}\text{F}_3$}}     & {{$R\bar{3}C~(167)$}}     & {{600}}     & {{$16\times16\times16$}} & {Cr:3} &{Cr:2.86}            \\ 
{27} &{{$\text{Mn}\text{O}$}}     & {{$ P6_3mc~(186)$}} & {{600}}     & {{$18\times18\times10$}} & {Mn:4} &{Mn:4.52}       \\
{28} &{{$\text{Ca}\text{Mn}\text{N}_2$}}     & {{$P6_3/mmc~(194)$}}     & {{600}}     & {{$18\times18\times6$}} & {Mn:4} &{Mn:3.39} \\
{29} &{{$\text{Ba}_2\text{Fe}\text{Ge}_2\text{O}_7$}}     & {{$P\overline{4}2_1m~(113)$}}     & {{600}}     & {{$6\times6\times8$}} & {Fe:5} &{Fe:3.69} \\
{30} &{{$\text{Ba}_2\text{Co}\text{Si}_2\text{O}_7$}}     & {{$P\overline{4}2_1m~(113)$}}     & {{600}}     & {{$6\times6\times8$}} & {Co:3} &{Co:2.69} \\
{31} &{{$\text{Sr}_2\text{Co}\text{Ge}_2\text{O}_{7}$}}   & {{$P\overline{4}2_1m~(113)$}}     & {{600}}     & {{$6\times6\times8$}} & {Co:3} &{Co:2.67} \\
{32} &{{$\text{V}\text{F}_4$}}                            & {{$P2_1/c~(14)$}}     & {{600}}     & {{$12\times6\times12$}} & {V:4} &{V:1.00} \\
{33} &{{$\text{Ca}_2\text{Co}\text{Te}\text{O}_{6}$}}     & {{$P2_1/c~(14)$}}     & {{600}}     & {{$8\times8\times6$}} & {Co:3} &{Co:2.69} \\
{34} &{{$\text{Ni}\text{F}_2$}}                           & {{$Pnnm~(58)$}}     & {{600}}     & {{$10\times10\times16$}} & {Ni:5} &{Ni:1.81} \\
{35} &{{$\text{Li}\text{Fe}_2\text{F}_{6}$}}              & {{$P4_2nm~(102)$}}     & {{600}}     & {{$10\times10\times6$}} & {Fe:5} &{Fe:4.088} \\
{36} &{{$\text{Fe}\text{H}\text{O}_2$}}                   & {{$P2_12_12_1~(19)$}}     & {{600}}     & {{$16\times12\times6$}} & {Fe:5} &{Fe:4.27}\\
{37} &{{$\text{Ca}\text{Mn}\text{O}_3$}}     & {{$Pnma~(62)$}}     & {{600}}     & {{$10\times8\times10$}} & {Mn:4} &{Mn:2.94}             \\
{38} &{{$\text{Ca}\text{V}\text{O}_3$}}     & {{$Pnma~(62)$}}     & {{600}}     & {{$10\times8\times10$}} & {V:4} &{V:0.95}             \\
{39} &{{$\text{La}\text{Fe}\text{O}_3$}}     & {{$Pnma~(62)$}}     & {{600}}     & {{$10\times10\times6$}} & {Fe:5} &{Fe:4.18}             \\
{40} &{{$\text{La}\text{V}\text{O}_3$}}     & {{$Pnma~(62)$}}     & {{600}}     & {{$10\times6\times10$}} & {V:4} &{V:1.85}            \\
{41} &{{$\text{Mn}\text{Se}\text{O}_4$}}     & {{$Pnma~(62)$}}     & {{600}}     & {{$10\times6\times8$}} & {Mn:4} &{Mn:4.63}             \\
{42} &{{$\text{Na}\text{Pr}_2\text{Os}\text{O}_6$}}     & {{$P2_1/c~(14)$}}     & {{600}}     & {{$10\times8\times6$}} & {Pr:7;Os:2} &{Pr:2.00;Os:2.16}             \\
{43} &{{$\text{Na}\text{Pr}_2\text{Ru}\text{O}_6$}}     & {{$P2_1/c~(14)$}}     & {{600}}     & {{$10\times8\times6$}} & {Pr:7;Ru:2} &{Pr:2.00;Ru:0.67}             \\
{44} &{{$\text{Nd}\text{Rh}\text{O}_3$}}     & {{$Pnma~(62)$}}     & {{600}}     & {{$10\times6\times10$}} & {Nd:7;Rh:2} &{Nd:3.00;Rh:0.00}             \\
{45} &{{$\text{Pr}\text{Ru}\text{O}_3$}}     & {{$Pnma~(62)$}}     & {{600}}     & {{$10\times6\times10$}} & {Pr:7;Ru:2} &{Pr:2.00;Ru:0.71}            \\
{46} &{{$\text{Sc}\text{V}\text{O}_3$}}     & {{$Pnma~(62)$}}     & {{600}}     & {{$10\times8\times10$}} & {V:4} &{V:1.83}             \\
{47} &{{$\text{Sm}\text{Rh}\text{O}_3$}}     & {{$Pnma~(62)$}}     & {{600}}     & {{$10\times6\times10$}} & {Sm:7;Rh:2} &{Sm:5.08;Rh:0.00}             \\
{48} &{{$\text{Ca}\text{La}\text{Cr}\text{Mo}\text{O}_6$}}     & {{$Pc~(7)$}}     & {{600}}     & {{$10\times8\times6$}} & {Cr:5;Mo:2} &{Cr:2.93;Mo:1.67}             \\
{49} &{{$\text{La}_2\text{Mn}\text{Rh}\text{O}_6$}}     & {{$P2_1/c~(14)$}}     & {{600}}     & {{$10\times8\times6$}} & {Mn:5;Rh:2} &{Mn:3.99;Rh:0.10}             \\
{50} &{{$\text{Li}\text{Fe}\text{F}_4$}}     & {{$P2_1/c~(14)$}}     & {{600}}     & {{$10\times10\times10$}} & {Fe:5} &{Fe:4.4}             \\

\bottomrule
\end{tabular}
\label{dft-details}
\end{table}
}

\footnotesize